# PhotoD with LSST: Stellar Photometric Distances Out to the Edge of the Galaxy

Lovro Palaversa[1], Željko Ivezić[2], Neven Caplar[2], Karlo Mrakovčić[3], Bob Abel[4,5], Oleksandra Razim[1], Filip Matković[6], Connor Yablonski[2], Toni Šarić[7], Tomislav Jurkić[3], Sandro Campos[8], Melissa DeLucchi[8], Derek Jones[2], Konstantin Malanchev[8], Alex I. Malz[8], Sean McGuire[8], and Mario Jurić[2]

[1] Ruđer Bošković Institute, Bijenička cesta 54, 10000 Zagreb, Croatia  
[2] DIRAC Institute and the Department of Astronomy, University of Washington, 3910 15th Avenue NE, Seattle, WA 98195, USA  
[3] Faculty of Physics, University of Rijeka, Radmile Matejčić 2, 51000 Rijeka, Croatia  
[4] Olympic College, 1600 Chester Avenue, Bremerton, WA 98337, USA  
[5] DiRAC Institute, University of Washington, Box 351580, Seattle, WA 98195, USA  
[6] Hvar Observatory, Faculty of Geodesy, University of Zagreb, Kačićeva 26, 10000 Zagreb, Croatia  
[7] University of Split—FESB, R. Boškovića 32, 21000 Split, Croatia  
[8] The McWilliams Center for Cosmology & Astrophysics, Department of Physics, Carnegie Mellon University, Pittsburgh, PA 15213, USA

*Received 2024 September 1; revised 2024 December 11; accepted 2024 December 27; published 2025 February 4*

## Abstract

As demonstrated with the Sloan Digital Sky Survey (SDSS), Pan-STARRS, and most recently with Gaia data, broadband near-UV to near-IR stellar photometry can be used to estimate distance, metallicity, and interstellar dust extinction along the line of sight for stars in the Galaxy. Anticipating photometric catalogs with tens of billions of stars from Rubin's Legacy Survey of Space and Time (LSST), we present a Bayesian model and pipeline that build on previous work and can handle LSST-sized datasets. Likelihood computations utilize MIST/Dartmouth isochrones and priors are derived from TRILEGAL-based simulated LSST catalogs from P. Dal Tio et al. The computation speed is about 10 ms per star on a single core for both optimized grid search and Markov Chain Monte Carlo methods; we show in a companion paper by K. Mrakovčić et al. how to utilize neural networks to accelerate this performance by up to an order of magnitude. We validate our pipeline, named PhotoD (in analogy with photo-$z$, photometric redshifts of galaxies) using both simulated catalogs and SDSS, DECam, and Gaia photometry. We intend to make LSST-based value-added PhotoD catalogs publicly available via the Rubin Science Platform with every LSST data release.

*Unified Astronomy Thesaurus concepts:* Distance measure (395); Distance indicators (394); Stellar distance (1595); Extinction (505); Interstellar extinction (841); Reddening law (1377)

## 1. Introduction

In order to map the Milky Way in three dimensions, distances to its stars must be accurately estimated. Kinematic studies based on proper motion data also require estimates of stellar distances. There are a variety of astronomical methods to estimate distances to stars, ranging from direct geometric (trigonometric) methods for nearby stars to indirect methods based on astrophysics for more distant stars.

As demonstrated with the Sloan Digital Sky Survey (SDSS; Ž. Ivezić et al. 2008; M. Jurić et al. 2008), Pan-STARRS (G. M. Green et al. 2014, 2019), and most recently with Gaia data (C. A. L. Bailer-Jones et al. 2021), broadband near-UV to near-IR stellar photometry is sufficient to estimate distance, metallicity, and interstellar dust extinction along the line of sight for stars in the Galaxy. In analogy with photo-$z$, the photometric redshifts of galaxies, hereafter we refer to these methods as photo-D. The photo-D method is conceptually quite simple: multidimensional color tracks (either empirical or model-based), parameterized by luminosity, metallicity, and extinction, are fit to observed colors and the best fit produces estimates of these three model parameters. The method relies on strong correlations between stellar colors and stellar luminosity for dominant stellar populations such as main-sequence stars, red giants, white dwarfs (WDs), and even for the majority of unresolved binary stars. These are the same correlations that are responsible for the abundant structure seen in the Hertzsprung–Russell diagram. In addition to colors and luminosity, these correlations also involve metallicity in the case of main-sequence and red giant stars, and surface gravity in the case of WDs. For the youngest main-sequence stars, stellar age may play a role, too. Measured stellar colors are also affected by interstellar dust extinction along the line of sight toward the star. Consequently, sufficiently accurate measurements of apparent brightness and sufficient number of UV to IR colors, such as those that Rubin Observatory's Legacy Survey of Space and Time (LSST; Ž. Ivezić et al. 2019) will provide, can be used to accurately estimate these parameters, and ultimately stellar distances.

LSST-based stellar distance estimates will significantly improve available distance catalogs, such as those recently produced by G. M. Green et al. (2019) and C. A. L. Bailer-Jones et al. (2021). First, the sample size will be increased by more than an order of magnitude, and exceed 10 billion stars. Distance accuracy for stars with sufficiently small photometric errors will be within the 5%–10% range, or about twice as accurate as for surveys lacking the UV $u$ band (which provides metallicity constraints). LSST-based stellar distances will reach about 10 times further than Gaia's color-based distances and will be transformative for studies of the Milky Way in general, and of its stellar and dark matter halos in particular.

In this paper we present a Bayesian model and pipeline that build on previous work and can handle LSST-sized datasets. In Section 2, we describe our methodology and in Section 3 we test the pipeline using both simulated catalogs and SDSS, DECam, and Gaia photometry. We discuss possibilities for further improvements and catalog public release plans in Section 4.

## 2. Methodology

In this section we discuss a Bayesian method for stellar photometric distance estimation and its implementation. We start with a brief overview of the Bayesian methodology and then discuss in detail our choices of likelihoods and priors, and how they differ from previous work. A pipeline implementation of this method and a discussions of its performance are presented in the next section.

### 2.1. Bayesian Approach to Stellar Photometric Distance Estimation

In the most general terms, our aim is to estimate for each star an array (a vector) of model parameters $\theta$, for some model $\mathcal{M}$ (e.g., main-sequence stars), using data vector $D$ and priors for $\mathcal{M}$ and $\theta$. Data, or observations, include multiband photometry that is used to construct colors, $c$. In the case of SDSS, colors include $u-g$, $g-r$, $r-i$, and $i-z$, and in the case of LSST also $z-y$. In addition to colors, $D$ also includes an apparent magnitude and hereafter we choose the $r$-band magnitude. Therefore, $D = (r, c)$. Observations also provide stellar sky coordinates and we address their role further below when discussing priors.

Models $\mathcal{M}$ (either empirical or computational, see below) need to provide stellar colors as functions of model parameters $\theta$. The three principal parameters that control stellar colors at a fixed stellar age and at the accuracy level relevant here (~1%) include absolute magnitude (here chosen in the $r$ band, $M_r$), metallicity ([Fe/H]), and surface gravity. In the case of main-sequence stars and red giants, the color tracks can be expressed as functions of $M_r$ and [Fe/H] along an isochrone, without having to explicitly specify surface gravity. With other populations, such as WDs, the roles of metallicity and surface gravity are different; we will assume here for notational simplicity that intrinsic stellar colors depend on $M_r$ and [Fe/H]. We do not consider stellar age as a model parameter and use it as a model label for reasons discussed below.

The observed stellar colors also depend on the interstellar dust extinction along the line of sight. Hereafter we will assume (and justify further below) that dust extinction is fully specified by a single model parameter, $A_r$. $A_r$ is extinction in the $r$ band and extinction in other bands is proportional to $A_r$, with known constants of proportionality (this assumption can be relaxed, see below). Once $M_r$ and $A_r$ are constrained, that is, the posterior probability distribution $q(Q_r)$, where $Q_r = M_r + A_r$ is known, the distance modulus $\Delta$ can be computed using relationship

$$r = M_r + A_r + \Delta. \quad (1)$$

In practice the uncertainty of the observed magnitude, $r^{\rm obs}$, is always much smaller than the width of the $q(Q_r)$ distribution (at the bright end by an order of magnitude, ~0.01 mag versus ~0.1 mag). Therefore

$$p(\Delta) = q(r^{\rm obs} - \Delta), \quad (2)$$

with the mean values related as $\bar{\Delta} = r^{\rm obs} - \overline{Q_r}$ and the uncertainty of $\Delta$ is approximately equal to the standard deviation of $q(Q_r)$. For related discussion, please see Section 2.4 in C. A. L. Bailer-Jones et al. (2021).

In summary, $D = (r, c)$ and $\theta = (M_r, [\text{Fe/H}], A_r)$. Data $D$ and model parameters $\theta$ are related via the Bayes theorem (see, e.g., Chapter 5 in Ž. Ivezić et al. 2020)

$$p(\mathcal{M}, \theta | D, I) = \frac{p(D | \mathcal{M}, \theta, I)\, p(\mathcal{M}, \theta | I)}{p(D | I)}, \quad (3)$$

where $I$ is prior information. Strictly speaking, the vector $\theta$ should be labeled by $\mathcal{M}$ since different models may be described by different parameters (e.g., main-sequence stars versus WDs).

The result $p(\mathcal{M}, \theta | D, I)$ is called the posterior probability density function (PDF) for model $\mathcal{M}$ and parameters $\theta$, given data $D$ and other prior information $I$. This term is a $(k+1)$-dimensional PDF in the space spanned by $k=3$ model parameters and the model index $\mathcal{M}$. The term $p(D | \mathcal{M}, \theta, I)$ is the likelihood of data given some model $\mathcal{M}$ and given some fixed values of parameters $\theta$ describing it, and all other prior information $I$. The term $p(\mathcal{M}, \theta | I)$ is the a priori joint probability, or simply prior, for model $\mathcal{M}$ and its parameters $\theta$ in the absence of any of the data used to compute likelihood. The prior can be expanded as

$$p(\mathcal{M}, \theta | I) = p(\theta | \mathcal{M}, I)\, p(\mathcal{M} | I). \quad (4)$$

The term $p(D | I)$ is the probability of data, or the prior predictive probability for $D$. It provides proper normalization for the posterior PDF; for simplicity, it is usually not explicitly computed when estimating model parameters: rather, $p(\mathcal{M}, \theta | D, I)$ for a given $\mathcal{M}$ is renormalized so that its integral over all model parameters $\theta$ is unity. The integral of the prior $p(\theta | \mathcal{M}, I)$ over all parameters should also be unity, but for the same reason, calculations of the posterior PDF are often done with an arbitrary normalization. An important exception is model selection discussed further below in Section 2.5, where the correct normalization of the product $p(D | \mathcal{M}, \theta, I)\, p(\theta | \mathcal{M}, I)$ is crucial.

Our approach adopted here is essentially the same as used in recent papers[9] by, e.g., C. A. L. Bailer-Jones (2011), G. M. Green et al. (2014, 2019), R. Lallement et al. (2014), K. D. Gordon et al. (2016), A. B. A. Queiroz et al. (2018), and C. A. L. Bailer-Jones et al. (2021). The main differences compared to these works include the following.

1. The use of multiple stellar populations, in addition to main-sequence stars and red giants (WDs, unresolved binary stars, and blue horizontal branch (BHB) stars; and potentially Miras, quasars, and RR Lyrae stars too, which can also be recognized and rejected using variability).
2. Improved color tracks for main-sequence stars and (especially) red giants, including the use of very young (<1 Gyr) populations and an extended [Fe/H] range.
3. Priors based on sophisticated TRILEGAL simulations by P. Dal Tio et al. (2022) that include multiple stellar

---

[9] This approach greatly simplifies when studying faint and distant blue halo stars, as in, e.g., M. Jurić et al. (2008). Such stars are beyond the dust layer, which is confined close to the disk and thus $A_r$ can be obtained from IR maps; they have halo metallicities ([Fe/H] ~ −1.5), and can be assumed dominated by main-sequence stars. As a result, a simple functional relationship, $M_r = f(g - i)$, or its generalized version that accounts for the shift of $M_r$ as a function of metallicity (Ž. Ivezić et al. 2008), can be used to estimate distance in a straightforward manner.

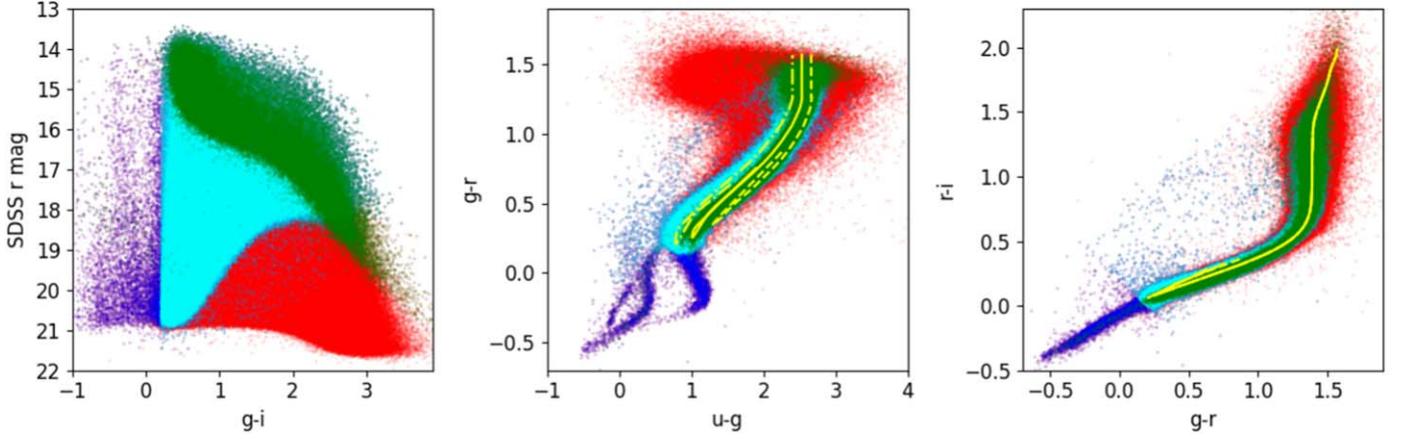

**Figure 1.** An illustration of multiple stellar populations. The left panel is a color–magnitude diagram for 841,000 stars from the SDSS Stripe 82 Standard Star Catalog (variable sources are excluded; K. Thanjavur et al. 2021) that have Gaia matches within 0″.15 (after correcting for proper motion using Gaia measurements). A subset of 415,000 stars with $r < 22$ and $u < 22$ are overplotted as blue dots, and 409,000 of those that also have $0.2 < g − i < 3.5$ (dominated by main-sequence stars and red giants) are overplotted as cyan dots. Finally, 63,000 stars that have a signal-to-noise ratio for Gaia's parallax measurements of at least 20 are shown as green dots (these stars can be used for the calibration of luminosity–color relations). The same symbol color scheme is used in the other two panels. The three yellow lines in the middle panel show the stellar locus parameterization used by G. M. Green et al. (2014) for three values of metallicity (left to right): [Fe/H] = −2, −1, and 0. In the right panel, the impact of metallicity on the color–color tracks is negligible and all three are indistinguishable from each other. In the bottom of the middle panel, at $-0.5 < g − r < 0$, the three dark blue sequences correspond to (from left to right) He WDs, H WDs, and BHB stars. The clouds of pale blue dots visible above the main stellar locus in the middle and right panels correspond to unresolved binary stars (V. Smolčić et al. 2004).

populations and also account for the Galaxy's bulge component (see Section 2.3 below for more details).

We discuss these improvements in detail in the next few sections.

### 2.2. Likelihood Computation

Given a chosen model (i.e., a stellar population) $\mathcal{M}$, the likelihood $p(\boldsymbol{D}|\mathcal{M}, \boldsymbol{\theta}, I)$ can be explicitly written as

$$\mathcal{L} \equiv p(\boldsymbol{D}|\mathcal{M}, \boldsymbol{\theta}, I) = p(\boldsymbol{c}|M_r, [\text{Fe/H}], A_r). \quad (5)$$

Assuming Gaussian photometric errors that are parameterized by a vector of color uncertainties $\boldsymbol{\sigma}$, the log-likelihood is given by

$$\ln(\mathcal{L}) = -\frac{N}{2}\ln(2\pi) - \sum_{i=1}^{N} \ln(\sigma_i) - \frac{1}{2}\sum_{i=1}^{N}\left(\frac{c_i^{\text{obs}} - c_i^{\text{mod}}}{\sigma_i}\right)^2, \quad (6)$$

where the summation is over all colors (for example, $N = 4$ for SDSS and $N = 5$ for LSST), where $c_i^{\text{obs}}$ are the observed colors and $c_i^{\text{mod}}$ are the model colors (they are functions of $M_r$, [Fe/H], and $A_r$ but for notational simplicity we do not explicitly list model parameters). Note that only the last sum involves model predictions for colors ($c_i^{\text{mod}}$).

The model colors can be computed as

$$\boldsymbol{c}^{\text{mod}} = \boldsymbol{c}_0(M_r, [\text{Fe/H}]) + \boldsymbol{\delta c}(A_r), \quad (7)$$

where $\vec{c}_0(M_r, [Fe/H])$ are intrinsic stellar colors for a given stellar population, and $\vec{\delta c}(A_r)$ are color corrections due to interstellar dust reddening. Equation (7) can be thought of as a set of three-dimensional data cubes, one for each color, that map the triplet ($M_r$, [Fe/H], $A_r$) to that color. The likelihood function can be thought of as a three-dimensional data cube that, for a given set of observed colors $c_i^{\text{obs}}$, maps the triplet ($M_r$, [Fe/H], $A_r$) to a one-dimensional scalar, that is, $\ln(\mathcal{L})$ is a scalar function of $M_r$, [Fe/H], and $A_r$.

The existence of multiple stellar populations in an SDSS photometric catalog is illustrated in Figure 1. With accurate multiband photometry that includes a UV band (here SDSS $u$), main-sequence stars, red giants, WDs, BHB stars, and unresolved binary stars can be reliably identified (UV photometry is also crucial for constraining metallicity). We discuss our choice of $\boldsymbol{c}_0(M_r, [\text{Fe/H}])$ for main-sequence stars and red giants next, and then for WDs, unresolved binary stars, and BHB stars.

#### 2.2.1. Empirical Luminosity–Color Tracks for Main-sequence Stars and Red Giants

Both M. Berry et al. (2012) and G. M. Green et al. (2014) used empirical color tracks for main-sequence stars and red giants (see the left panel in Figure 2, modeled after Figure 1 in G. M. Green et al. 2014) derived from SDSS data for globular clusters (for technical details, see Appendix A in Ž. Ivezić et al. 2008). These color tracks suffer from three problems. First, as can be seen in Figure 11 from G. M. Green et al. (2014), their predicted colors for subgiant stars between the main-sequence turnoff and red giant branch are too blue by about 0.1–0.2 mag. Second, their metallicity grid does not extend to the [Fe/H] > 0 range relevant for some disk stars. Finally, they correspond to very old populations (older than a few gigayears) and cannot be used for stars younger than about 1–2 Gyr.

We use a combination of SDSS and Gaia data to demonstrate the first problem. The middle panel in Figure 2 shows a clear discrepancy between SDSS-based empirical tracks and data for subgiant stars, using parallax-based absolute magnitudes. Nevertheless, it is noteworthy that for main-sequence stars the agreement is excellent. Furthermore, the right panel in Figure 2 demonstrates that SDSS-based photometric distances and Gaia-based "photogeometric" distances from C. A. L. Bailer-Jones et al. (2021) for main-sequence stars are on the "same scale." We discuss these distance scales in more detail in Section 3.2.2.

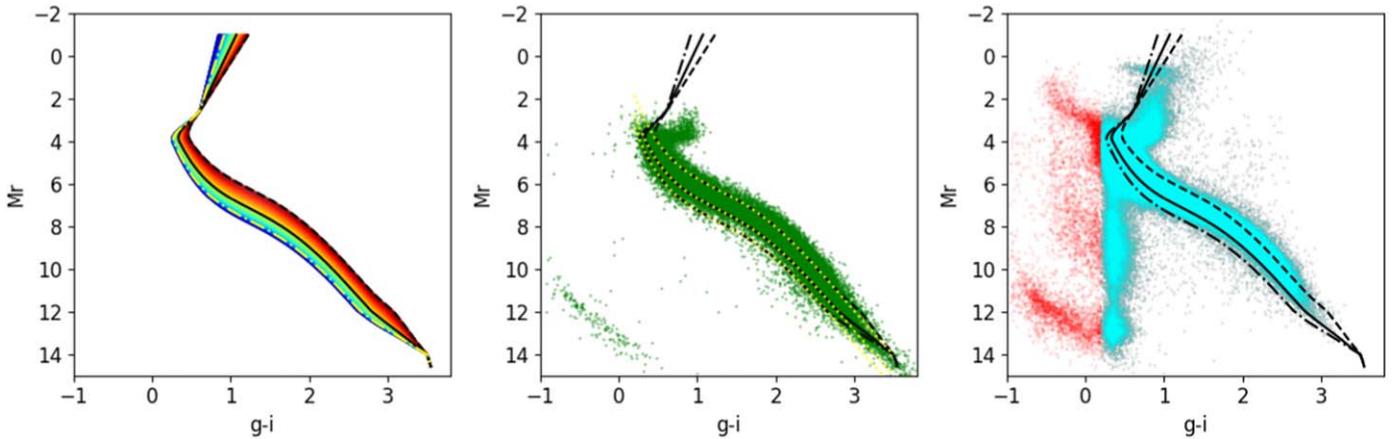

**Figure 2.** The left panel shows SDSS-based empirical absolute magnitude vs. color parameterization for main-sequence stars and red giants. The data are color coded by metallicity, ranging from [Fe/H] = −2.5 to 0 (blue to red). The three lines correspond to three values of metallicity: [Fe/H] = −2, −1, and 0 (dotted–dashed, solid, and dashed, respectively). The middle panel shows a sample of 63,000 stars that have a signal-to-noise ratio for Gaia's parallax measurements of at least 20 (WDs can be seen in the lower left corner). Their absolute magnitudes are derived from parallax measurements. The dotted–dashed, solid, and dashed black lines are the same as in the left panel. For comparison, the essentially identical yellow dotted lines were computed using Equations (A2) and (A7) from Ž. Ivezić et al. (2008). Note the discrepancy between these parameterizations and data for subgiant stars ($M_r$ ~ 3−4 and $g − i$ ~ 0.8−1.1). The right panel shows a sample of 415,000 stars with $r < 22$ and $u < 22$ as red dots (shown by blue dots in Figure 1), and 409,000 of those that also have $0.2 < g − i < 3.5$ as cyan dots. Their absolute magnitudes were computed using the so-called "photogeometric" distances from C. A. L. Bailer-Jones et al. (2021). The dotted–dashed, solid, and dashed black lines are the same as in the left and middle panels. About 10,000 stars below the main sequence (about 2.5% of the full sample) seen at $g − i = 0.4$ and $M_r > 7$ are predominantly found at the faint end ($r > 20$) and may be outliers due to the low photometric signal-to-noise ratio.

We address all three problems by augmenting the SDSS-based empirical isochrones that correspond to old populations with model-based isochrones that span a range of ages and a wider range of metallicities, which results in better agreement with the observations.

### 2.2.2. Model-based Isochrones for Main-sequence Stars and Red Giants

We considered two sets of isochrones: the Dartmouth Stellar Evolution Database[10] (DSED; A. Dotter et al. 2008), and PARSEC[11] (A. Bressan et al. 2012) isochrones. Both isochrone sets span adequate ranges of age and metallicity. Unfortunately, the computed color sequences show discrepancies with the observed SDSS stellar locus at the level of 0.1–0.2 mag and cannot be used without further adjustments.

We "augment" the empirical SDSS-based isochrones in two steps.

1. We extend its metallicity range from [Fe/H] = 0 to [Fe/H] = +0.5 by linear extrapolation of the color versus metallicity dependence around [Fe/H] = 0. We adjust the gradient by a multiplicative factor (0.7) to ensure that the bright end of the stellar locus at [Fe/H] = +0.5 agrees with SDSS–Gaia data (shown in the middle panel in Figure 2).
2. For $M_r < M_{TO}$, where turnoff absolute magnitude $M_{TO}$ depends on age and ranges from $M_{TO} = 4$ for age = 1 Gyr to $M_{TO} = 5$ for age = 10 Gyr, we "attach" model-based isochrones (we use DSED isochrones hereafter) to the empirical SDSS-based isochrones.

Examples of the resulting color tracks for two representative ages are shown in Figure 3. Subgiant stars that could not be fit with empirical SDSS-based tracks can now be explained with tracks for old stars and intermediate-range metallicity. We have computed such tracks for a grid of ages but believe that the two choices shown in Figure 3 (1 and 10 Gyr) should suffice for "nonspecialized" bulk processing. The reason is that the loci for ages above 1–2 Gyr look very similar to each other, while the fraction of stars younger than 1 Gyr is very small at the faint apparent magnitude levels probed by SDSS and LSST. Of course, in sky regions with intensive star formation, a "specialized" approach with a fine age grid can be easily executed.

We note strong degeneracies in color space between giants and main-sequence stars (see Figure 4). They are especially strong for subgiant stars with $2 < M_r < 4$, for which it is always possible to find a matching main-sequence star closer in four-dimensional color distance than 0.02 mag (and for $3 < M_r < 4$ even closer than 0.01 mag). Due to photometric scatter, such subgiant stars can be misidentified as a main-sequence star and absolute magnitude errors can range up to several magnitudes.

These color tracks for main-sequence stars, BHB stars, and red giants account for an overwhelming majority of stars expected in SDSS and LSST catalogs (approximately >95% but the fraction varies with apparent magnitude and sky position). Nevertheless, we also explicitly account for a few additional populations: WDs and unresolved binary stars.

### 2.2.3. Luminosity–Color Tracks for White Dwarfs

High-precision SDSS photometry clearly shows two WD sequences in the $g − r$ versus $u − g$ color–color diagram (see, e.g., Figures 23 and 24 in Ž. Ivezić et al. 2007; as well as the middle panel in Figure 1). A comparison with models from P. Bergeron et al. (1995), as well as with SDSS spectra, reveals that the two sequences correspond to H and He WDs, with the mean $\log(g) = 8.0$ for the H sequence and $\log(g) = 8.5$ for the He sequence. Upper limits for the scatter of $\log(g)$ around these mean sequences appear as no more than 0.5. We use three modern WD catalogs: the Montreal White Dwarf Database[12] (P. Dufour et al. 2017), the Gaia EDR3

---

[10] See http://stellar.dartmouth.edu/models.
[11] http://stev.oapd.inaf.it/cgi-bin/cmd_3.7
[12] https://www.montrealwhitedwarfdatabase.org

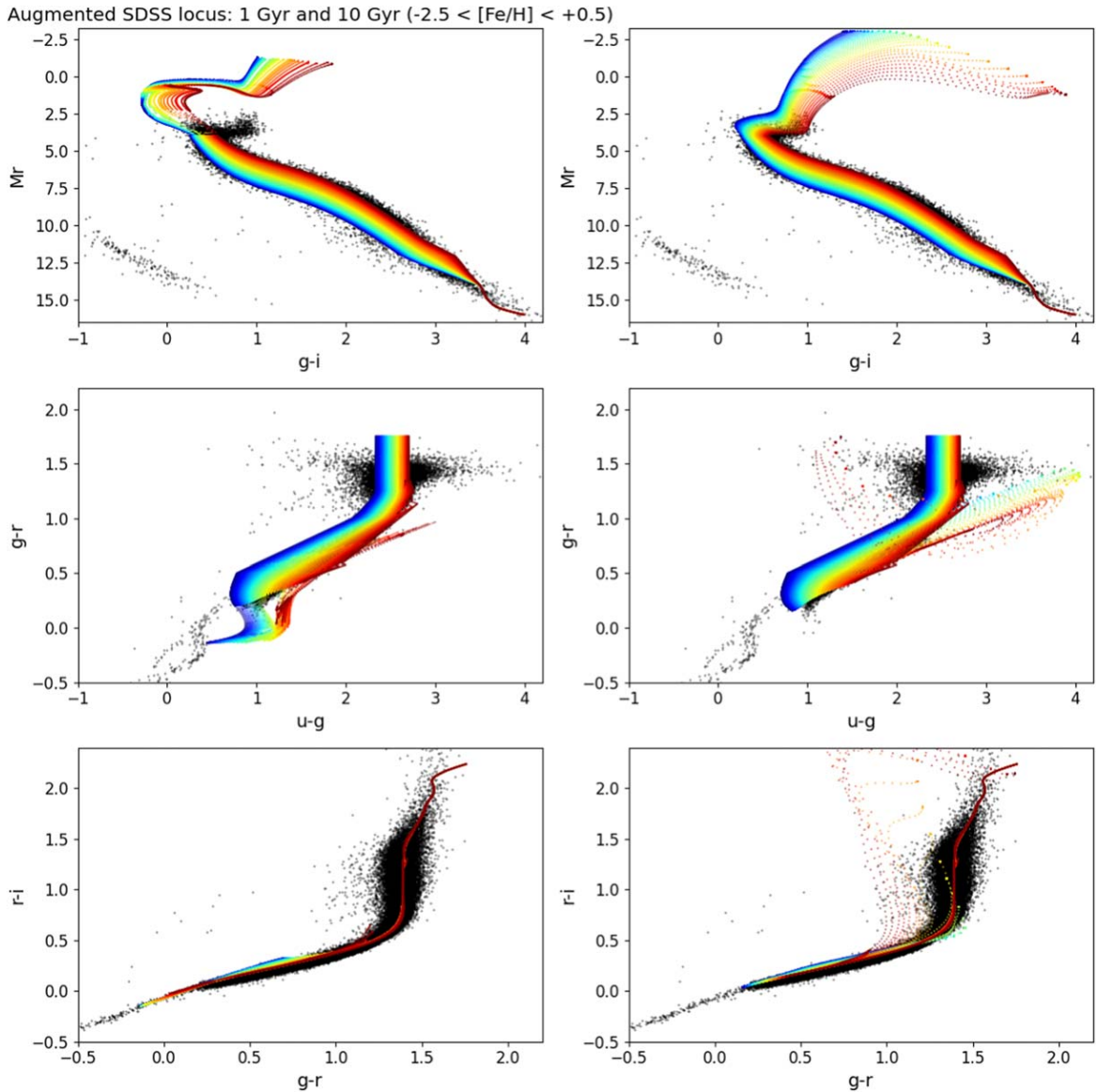

**Figure 3.** Augmented SDSS color tracks for two isochrone ages (left: 1 Gyr; right: 10 Gyr) and the full metallicity range ($-2.5 < $ [Fe/H] $< +0.5$, color coded linearly from blue to red). The black dots show the same sample of 63,000 stars from the middle panel in Figure 2. As can be seen in the top right panel, subgiant stars that could not be fit with empirical SDSS-based tracks (see the middle panel in Figure 2) can be explained with tracks for old stars and intermediate-range metallicity. The sharp feature protruding from the main locus in the middle two panels corresponds to most luminous and evolved high-metallicity stars. Note that diagrams in the bottom row, which do not include the $u$ band, show very little dependence on metallicity.

White Dwarf Catalog[13] (N. P. Gentile Fusillo et al. 2021), and the Gaia-based White Dwarf Database[14] (E. M. Garcìa-Zamora et al. 2023) to validate the P. Bergeron et al. (1995) models and derive small color offsets that bring the models in perfect agreement with these three datasets.

We consider only WDs with a cataloged DA, DB, or DC spectral class, at Galactic latitudes further than $10°$ from the plane, with SDSS photometry, apparent magnitudes $5 \leqslant m \leqslant 22$ in any band $m$, and $15 \leqslant r \leqslant 19$, where $r$ is the SDSS $r$-band magnitude. Observed magnitudes are corrected for interstellar dust extinction using maps from D. J. Schlegel et al. (1998) and per-band extinction coefficients discussed in Section 2.2.6 below (for the nearest WDs in the sample, this correction may be somewhat overestimated). We group the DB and DC spectral classes as He dwarfs (1939 objects), while the DA spectral class corresponds to H WDs (9307 objects). Their absolute magnitudes were calculated using "photogeometric" distances from C. A. L. Bailer-Jones et al. (2021).

For each WD type (H and He) and SDSS color, we bin the data into 20 $M_r$ bins and compute the median value of a given color in each bin. The color versus $M_r$ sequences from P. Bergeron et al. (1995) are then slightly (up to 0.1 mag) shifted in color so that the mean offset for all bins vanishes. The only case where a small (up to 0.1 mag) linear adjustment of the model track was needed is the $u - g$ color for H models at $g - r < -0.3$. After these color adjustments are applied, all color versus $M_r$ model sequences were linearly interpolated to a common $M_r$ grid ($8.5 < M_r < 14.5$, with a step of 0.02 mag). The resulting two model tracks for H and He WDs are shown in Figure 5.

---

[13] https://warwick.ac.uk/fac/sci/physics/research/astro/research/catalogues/
[14] https://cdsarc.cds.unistra.fr/viz-bin/cat/J/A+A/679/A127

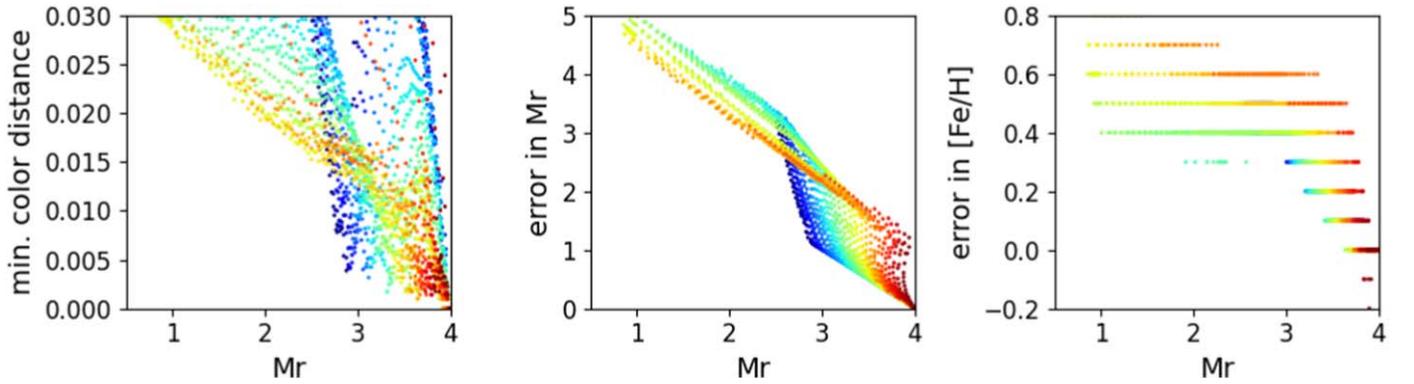

**Figure 4.** Analysis of color degeneracies between giants and main-sequence stars. The left panel shows minimum color distance in four-dimensional color space ($u − g$, $g − r$, $r − i$, and $i − z$) between a position on the locus with $M_r < 4$ (giants) and any position on the main-sequence $M_r > 4$ locus (in magnitudes). Symbols are color coded by metallicity, linearly from blue to red for the range −2.5 to 0.5. The middle and right panels show errors in absolute magnitude and metallicity when a giant is misidentified as a main-sequence star closest to it in color space.

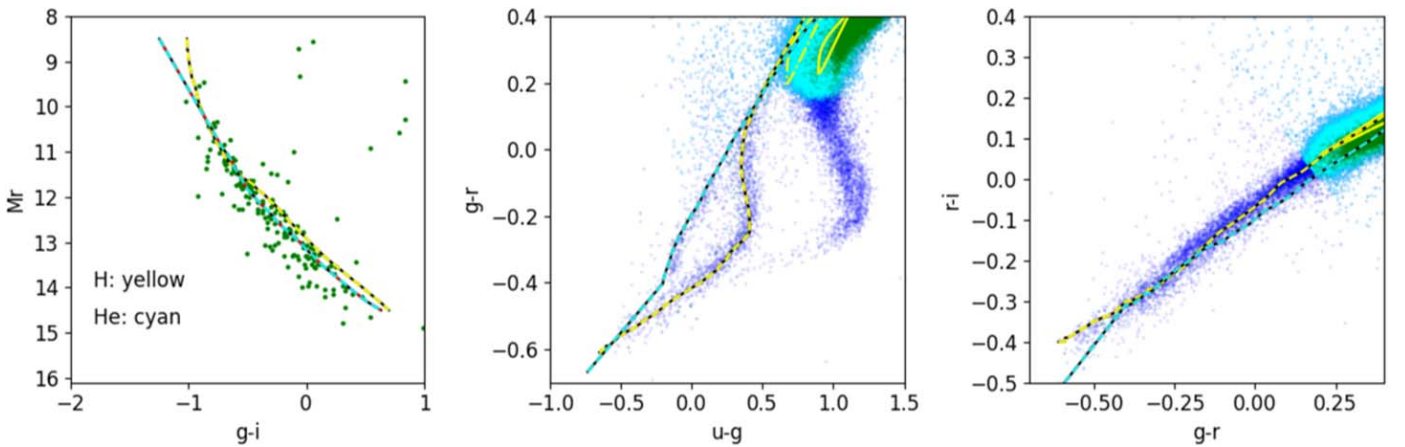

**Figure 5.** Luminosity–color tracks for H and He WDs based on models from P. Bergeron et al. (1995), with colors slightly shifted to bring them in agreement with the SDSS photometry. Data, shown as symbols, are the same as in Figure 1 and in the middle panel in Figure 2. Note the crucial role of the $u − g$ color for distinguishing H and He tracks.

### 2.2.4. Luminosity–Color Tracks for Unresolved Binaries

Following V. Smolčić et al. (2004), who discovered the so-called "second stellar locus" of unresolved binary stars in the SDSS dataset, we generate luminosity–color tracks for unresolved binaries consisting of an M dwarf and a WD. We limit models to M dwarfs because WDs would have a negligible impact on colors of more luminous main-sequence stars. M dwarfs are parameterized with $M_r$ and [Fe/H], and H/He WDs with $M_r$. Therefore, model tracks for unresolved binary stars would have three model parameters. After numerical experimentation that revealed low model sensitivity to metallicity, we decided to consider only M dwarfs with two metallicity values corresponding to mean metallicity for disk and halo populations: [Fe/H] = 0 and [Fe/H] = −1.5, respectively. For the remaining two model parameters, we selected $M_r$ corresponding to the total system luminosity and the component luminosity ratio in the $r$ band.

We sample the M dwarf luminosity in the range $8.5 \leqslant M_r \leqslant 14.5$, with a step of 0.1 mag, and for each value generate a track by adding the luminosity of a WD, where the track is sampled on the same grid of $M_r$ but using WD models (separately for H and He models). There are four model families (for two families of WD models and two values of [Fe/H] for M dwarfs), each with 3721 $M_r$ versus color entries. The color tracks for H WDs and [Fe/H] = −1.5 M dwarfs are shown in Figure 6. Analogous tracks for binaries composed of an M dwarf with [Fe/H] = 0 and either an H or He WD look very similar though not identical (differences are at most a few tenths of a magnitude). In practice, it will be nearly impossible to constrain M dwarf metallicity, and very hard to distinguish H and He WDs. It may turn out that it will be sufficient to use a single model family to fit LSST data (e.g., a combination of an H WD and an [Fe/H] = −1.5 M dwarf).

### 2.2.5. Luminosity–Color Tracks for Unaccounted Populations

There will always be sources that cannot be fully explained with any of the available families of luminosity–color tracks. For example, quasars and RR Lyrae are not considered here but can be easily recognized and removed from the sample in a straightforward manner since they are variable (e.g., see B. Sesar et al. 2007). The remaining nonvariable sources that are not well fit with available models can be recognized as "bad fits" (see the last subsection on model selection below).

There will be impostors, too, with measured colors consistent with available model(s), but with a very different nature (e.g., ultracold WDs "hiding" in the main stellar locus). As the time-domain information is built up with the progress of LSST, it might be possible to recognize them, for example, as high proper motion sources or perhaps using a broader wavelength range through cross correlation with other surveys (e.g., the Wide-field Infrared Survey Explorer or Roman surveys). While they will require specialized studies, based on TRILEGAL simulations we expect that their fraction in the sample should be very small (most likely <1%).

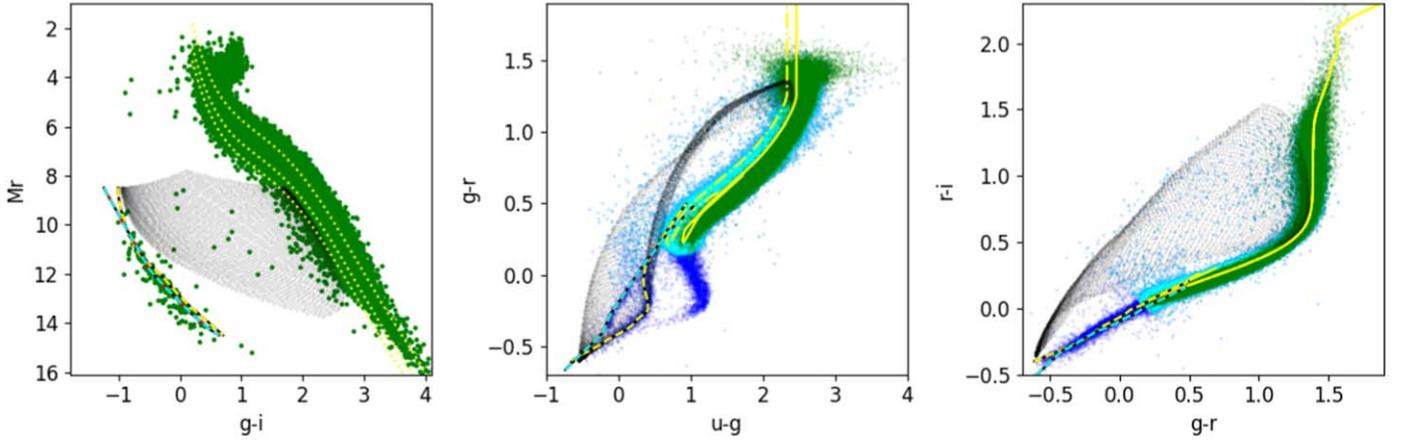

**Figure 6.** Luminosity–color tracks, shown with gray symbols, for unresolved binaries composed of an M dwarf with [Fe/H] = −1.5 and an H WD. Analogous tracks for binaries composed of an M dwarf with [Fe/H] = 0 and either an H or He WD look very similar though not identical. Data, shown as colored symbols, are the same as in Figure 1 and in the middle panel in Figure 2.

### 2.2.6. Accounting for Interstellar Dust Extinction

Given a stellar population and resulting color tracks for intrinsic unextincted colors, colors need to be corrected for interstellar dust extinction using an adequate dust extinction model. Additive color corrections $\delta c(A_r)$ (see Equation (7)) can be computed as

$$\delta c = (C_{m2} - C_{m1})\, A_r, \qquad (8)$$

where $m1$ and $m2$ stand for two bandpasses that define the color (e.g., $u$ and $g$). Dust extinction models, such as J. A. Cardelli et al. (1989) and E. L. Fitzpatrick (1999), parameterize extinction coefficients $C_m$ as functions of $R_V$, where $R_V = A_V/E(B-V)$ and $E(B-V)$ is the stellar "color excess" (J. A. Cardelli et al. 1989).

Studies, such as M. Berry et al. (2012), based on SDSS data (including the Galactic plane), find that the scatter of $R_V$ around its mean value $R_V = 3.1$ is very small. For this reason, we adopt empirical results from their Table 1: $C_m = (1.810, 1.400, 0.759, 0.561)$ in ($u, g, i, z$), respectively ($C_r = 1$ by definition). This choice does not preclude the use of our framework to fit for $R_V$ as the fourth free model parameter in Galactic plane regions with very large $A_r$ ($R_V$ is poorly constrained when $A_r$ is small); we can simply add multiple models with extincted colors generated using different values of $C_m$.

In our implementation, constraints on $A_r$ for two nearby stars, even if they have similar distances, are independent. We note that one could use the so-called hierarchical Bayesian modeling and specify the prior for the line-of-sight extinction profile: nearby stars would then jointly constrain it. For more details, see G. M. Green et al. (2014).

Finally, it is noteworthy to point out that the dust extinction vector is nearly parallel to the main stellar locus, and this fact may cause model parameter degeneracies for certain choices of colors (e.g., see Figure 9 in M. Berry et al. 2012). Such degeneracies are broken when multiple colors that span a wide wavelength range and extend into near-IR (such as the $i - z$ color) are available. For more details, see Section 2.8 in M. Berry et al. (2012).

### 2.3. TRILEGAL-based Priors

In addition to specifying the likelihood $p(\mathbf{c}|M_r, [\mathrm{Fe/H}], A_r)$ for a given model $\mathcal{M}$, we need to specify the prior probability distribution for model parameters, $p(M_r, [\mathrm{Fe/H}], A_r)$ (see Equation (4); we address the model probability $p(\mathcal{M}|I)$ further below). First, we assume that the stellar model parameters are unrelated to the distribution of interstellar dust along the line of sight

$$p(M_r, [\mathrm{Fe/H}], A_r) = p(M_r, [\mathrm{Fe/H}])\, p(A_r). \qquad (9)$$

We adopt a uniform prior for $p(A_r)$ using the values $A_r^{\mathrm{SFD}}$ taken from the D. J. Schlegel et al. (1998) dust extinction maps. To account for potential map errors, we set the maximum allowed value of $A_r$ as

$$A_r^{\max} = a\, A_r^{\mathrm{SFD}} + b, \qquad (10)$$

and $a = 1.3$ and $b = 0.1$, where the choice of these parameters allows for plausible upper limits for additive and multiplicative errors in the dust extinction maps. Therefore, $p(A_r) = 1/A_r^{\max}$ for $0 \leqslant A_r \leqslant A_r^{\max}$ and $p(A_r) = 0$ for $A_r > A_r^{\max}$.

The prior distribution of stellar model parameters, $p(M_r, [\mathrm{Fe/H}])$, depends on sky position and apparent magnitude (here $r$) due to the complex structure of the Milky Way. For example, G. M. Green et al. (2014) generated priors using SDSS-based analytic descriptions of the three-dimensional stellar distribution in the Milky Way (M. Jurić et al. 2008) and metallicity distribution for disk and halo components (Ž. Ivezić et al. 2008). Since LSST aims to provide good coverage of the Galactic plane and the bulge, where SDSS data did not provide strong constraints, we opted to utilize recent simulations that implement constraints from a variety of modern surveys (some include the Galactic plane and bulge regions). An analogous approach was taken by C. A. L. Bailer-Jones et al. (2021).

P. Dal Tio et al. (2022) have generated a mock catalog of Milky Way stars to LSST depth ($r = 27.5$) and over the entire LSST survey area. The simulation is based on the TRILEGAL code, incorporates all principal stellar populations, includes about 10 billion stars, and the catalog is publicly accessible through the NOIRLab Astro Data Lab.[15] We have developed code to query this mock LSST catalog using the Astro Data

---
[15] Available at https://datalab.noirlab.edu/.

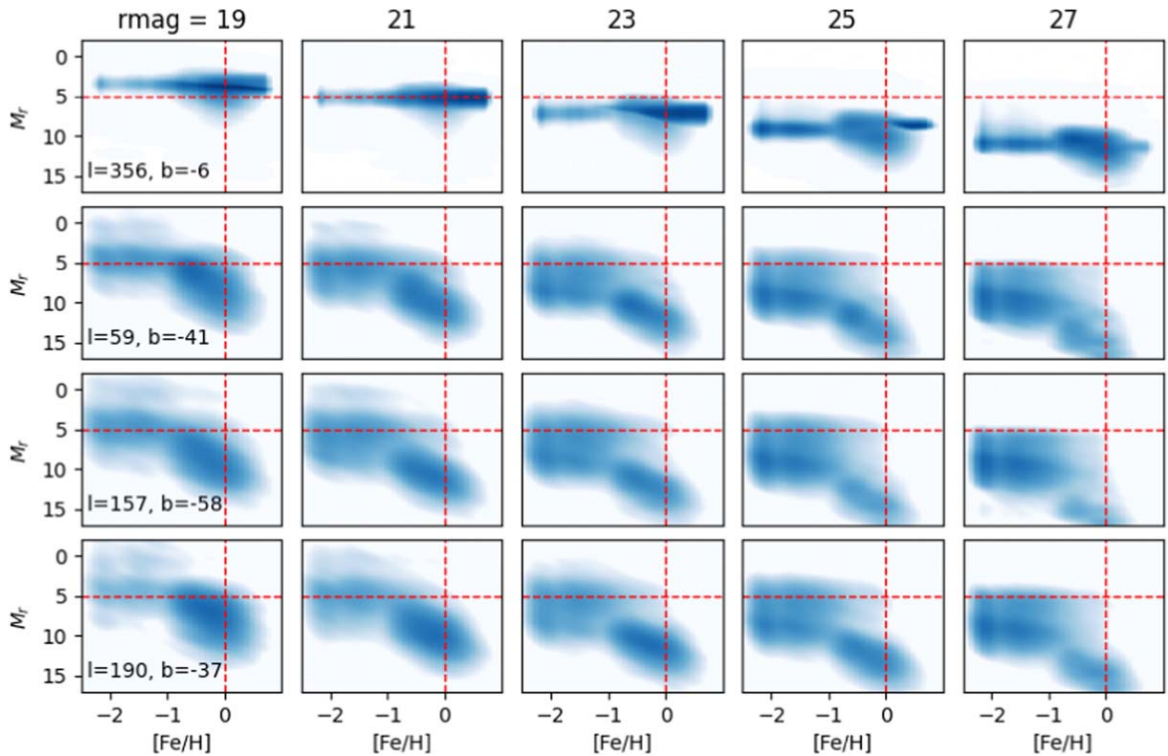

**Figure 7.** Examples of priors for model parameters, $p(M_r, [Fe/H])$, for main-sequence stars and red giants, for four sky positions (rows, for locations see the first column) and for several brightness levels in the LSST magnitude range (left to right: bright to faint). The integral of each map over $M_r$ and $[Fe/H]$ is unity. The dashed lines are the same in each panel and are added to guide the eye.

Lab portal. For a given position on the sky, we extract all catalog entries from an area of ~10 deg² (the size of the LSST camera's field of view) around it, then bin the sample by apparent $r$-band magnitude (27 bin centers from $r = 14$ to $r = 27$, with a bin width of 1 mag). Given an $r$-band-selected subsample, we separate all modeled populations and then bin them using adequate model parameters (e.g., $M_r$ and $[Fe/H]$ in the case of main-sequence stars and red giants). The resulting two-dimensional map is renormalized so that its integral over all model parameters is unity.

Figure 7 shows examples of such $p(M_r, [Fe/H])$ maps for a grid of apparent magnitudes and for several characteristic sky positions. Several features are noticeable: (i) there are two "clouds," at low and high metallicity, that correspond to halo and disk stars; (ii) the distribution toward the Galactic center (top row) is much more compact in the $M_r$ direction (because these subsamples are dominated by the bulge stars at similar distances); and (iii) the whole distribution shifts to fainter levels as the bin center ($r$) becomes fainter. We have experimented with different region sizes[16] and found that ~10 deg² (HEALPix $N_{side} = 16$ and 32, with 3072 and 12,288 pixels over the full sky) is sufficiently small to capture the variation of $p(M_r, [Fe/H])$ maps across the sky.

### 2.3.1. Mitigating the Specific Choice of Priors

The full Bayesian computation is an optimal method for estimating the best model parameters for individual stars. This method incorporates what is already known about the Milky Way structure through TRILEGAL-based priors. To use the same dataset for estimating the parameters of an alternative Galactic model, the likelihood function for stellar parameters should be reported instead of their Bayesian posterior probability distribution. For example, the latter approach was adopted by M. Berry et al. (2012). In order to enable both use cases, we intend to persist and report both the likelihood function and the Bayesian posterior probability distribution of the model parameters.

### 2.4. An Example of Bayesian Model Parameter Estimation

We use the isochrones shown in the right panels of Figure 3 and the priors shown in the second row of Figure 7 to illustrate the Bayesian model parameter estimation. Figure 8 shows the prior, likelihood, and posterior for a main-sequence star close to the turnoff point. Note how the prior, although much wider than the likelihood map, helps break the degeneracy[17] in the likelihood map (the two "islands").

Figure 9 shows two-parameter covariances and marginal distributions for this case. Note that true values are recovered within the expected uncertainties, as well as nonvanishing covariances between the parameters. The marginal distributions produced with the prior, likelihood, and posterior maps, shown in Figure 10, illustrate the improvement in the "knowledge" of model parameters between the prior and posterior, brought by color measurements via the likelihood map.

In the next section, we discuss a fast numerical pipeline implementation that can perform this computation for LSST-

---
[16] Following C. A. L. Bailer-Jones et al. (2021), we use HEALPix geometry, see https://healpix.jpl.nasa.gov.

[17] Priors can break degeneracies between the giant and dwarf stars because luminous stars become strongly disfavored at faint magnitudes (because an apparently faint giant star would imply a very large distance, beyond the presumed edge of the Galaxy at ~100 kpc; for example a giant star with $M_r = 0$ and $r = 22$ would imply a distance of ~250 kpc).

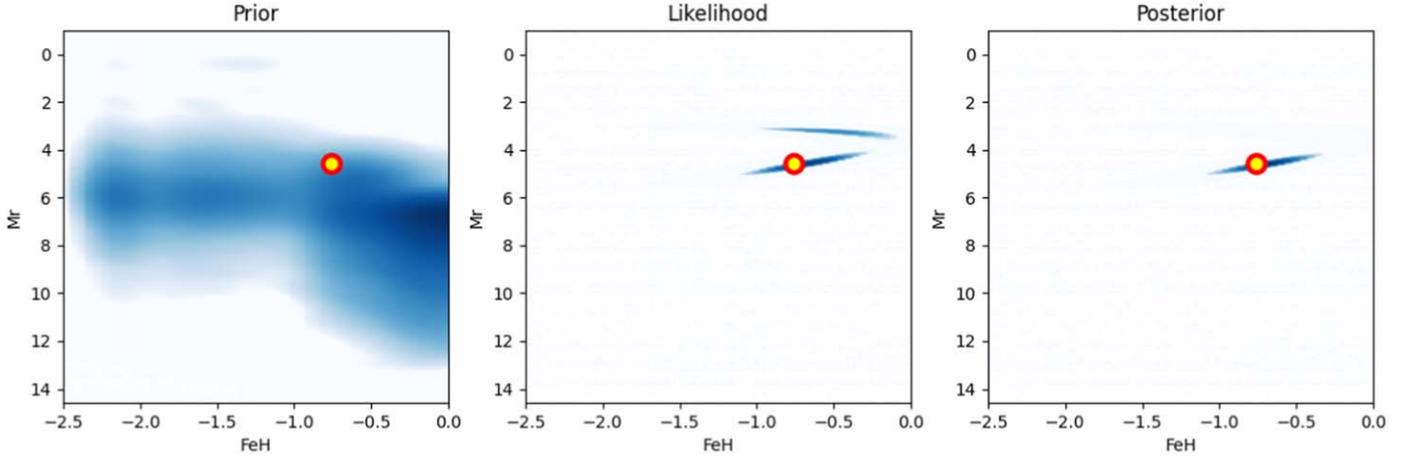

**Figure 8.** Maps of the prior (left), likelihood (middle; computed using Equation (6)), and posterior (right; computed using Equation (3)) for a simulated main-sequence star with $r = 21.2$, $(g − r) = 0.48$, $M_r = 4.58$, [Fe/H] = −0.76, and $A_r = 0.37$. The prior in the $A_r$ direction is uniform. The circles mark the values of the input $M_r$ and [Fe/H]. The posterior marginal distributions for all three model parameters are shown in Figure 9.

sized catalogs and provide a more quantitative analysis of the method's performance. We conclude this section with a brief discussion of the Bayesian model selection method for assigning posterior probabilities to each stellar population used to interpret observations.

### 2.5. Bayesian Model Selection

For each star and for each stellar population (that is, a model for color–magnitude tracks), a posterior three-dimensional data cube is produced. These posteriors are used when choosing the best model, as follows.

Bayes theorem as introduced by Equation (3) quantifies the posterior PDF of parameters describing a single model, with that model assumed to be true. To find out which of two models, say $\mathcal{M}_1$ and $\mathcal{M}_2$, is better supported by data, we compare their posterior probabilities via the odds ratio in favor of model $\mathcal{M}_2$ over model $\mathcal{M}_1$ as

$$O_{21} \equiv \frac{p(\mathcal{M}_2|\boldsymbol{D}, I)}{p(\mathcal{M}_1|\boldsymbol{D}, I)}. \quad (11)$$

The posterior probability for model $\mathcal{M}$ ($\mathcal{M}_1$ or $\mathcal{M}_2$) given data $\boldsymbol{D}$ and $p(\mathcal{M}|\boldsymbol{D}, I)$ in this expression, can be obtained from the posterior PDF $p(\mathcal{M}, \boldsymbol{\theta}|\boldsymbol{D}, I)$ in Equation (3) using marginalization (integration) over the model parameter space spanned by $\boldsymbol{\theta}$. The posterior probability that the model $\mathcal{M}$ is correct given data $\boldsymbol{D}$ (a number between 0 and 1) can be derived using Equations (3) and (4) as

$$p(\mathcal{M}|\boldsymbol{D}, I) = \frac{p(\boldsymbol{D}|\mathcal{M}, I) \, p(\mathcal{M}|I)}{p(\boldsymbol{D}|I)}, \quad (12)$$

where

$$E(\mathcal{M}) \equiv p(\boldsymbol{D}|\mathcal{M}, I) = \int p(\boldsymbol{D}|\mathcal{M}, \boldsymbol{\theta}, I) \, p(\boldsymbol{\theta}|\mathcal{M}, I) \, d\boldsymbol{\theta}, \quad (13)$$

is called the marginal likelihood (or the evidence) for model $\mathcal{M}$ and it quantifies the probability that the data $\boldsymbol{D}$ would be observed if the model $\mathcal{M}$ were the correct model. Since the marginal likelihood $E(\mathcal{M})$ involves integration of the data likelihood $p(\boldsymbol{D}|\mathcal{M}, \boldsymbol{\theta}, I)$, it is also called the global likelihood for model $\mathcal{M}$.

The global likelihood is a weighted average of the likelihood function, with the prior for model parameters acting as the weighting function. Alternatively, $E(\mathcal{M})$ is simply the integral over allowed parameter space of the posterior PDF before its renormalization to set this integral to unity (e.g., the integral of the posterior shown in the right panel in Figure 8). If the chosen model color tracks cannot explain the observed colors, the likelihood (Equation (6)) will never be very high and the resulting $E(\mathcal{M})$ will be low. We note that in the limit of a Gaussian posterior and flat priors, Bayesian evidence-based model selection becomes equivalent to $\chi^2$ selection from the frequentist statistical framework.

The probability of data, $p(\boldsymbol{D}|I)$, cancels out when the odds ratio is considered

$$O_{21} = \frac{E(\mathcal{M}_2) \, p(\mathcal{M}_2|I)}{E(\mathcal{M}_1) \, p(\mathcal{M}_1|I)} = B_{21} \frac{p(\mathcal{M}_2|I)}{p(\mathcal{M}_1|I)}. \quad (14)$$

In practice, the values of the odds ratio are interpreted using Jeffreys' scale (see, e.g., Chapter 5 in Ž. Ivezić et al. 2020); in particular, $O_{21} > 10$ represents "strong" evidence in favor of $\mathcal{M}_2$ ($\mathcal{M}_2$ is 10 times more probable than $\mathcal{M}_1$).

The ratio of global likelihoods, $B_{21} \equiv E(\mathcal{M}_2)/E(\mathcal{M}_1)$, is called the Bayes factor, and is equal to

$$B_{21} = \frac{\int p(\boldsymbol{D}|\mathcal{M}_2, \boldsymbol{\theta}_2, I) \, p(\boldsymbol{\theta}_2|\mathcal{M}_2, I) \, d\boldsymbol{\theta}_2}{\int p(\boldsymbol{D}|\mathcal{M}_1, \boldsymbol{\theta}_1, I) \, p(\boldsymbol{\theta}_1|\mathcal{M}_1, I) \, d\boldsymbol{\theta}_1}. \quad (15)$$

The vectors of parameters, $\boldsymbol{\theta}_1$ and $\boldsymbol{\theta}_2$, are explicitly indexed to emphasize that the two models may span vastly different parameter spaces (including the number of parameters per model).

The prior model probabilities, $p(\mathcal{M}_1|I)$ and $p(\mathcal{M}_2|I)$, are determined using the TRILEGAL-simulated catalog. For example, if model 1 is main-sequence stars and model 2 is WDs, we estimate the $p(\mathcal{M}_2|I)/p(\mathcal{M}_1|I)$ ratio by simply counting main-sequence stars and WDs in the corresponding $r$-magnitude bin.

In the case of $N_{\mathcal{M}}$ models, we assume that they represent an exhaustive model set and estimate the posterior probability of

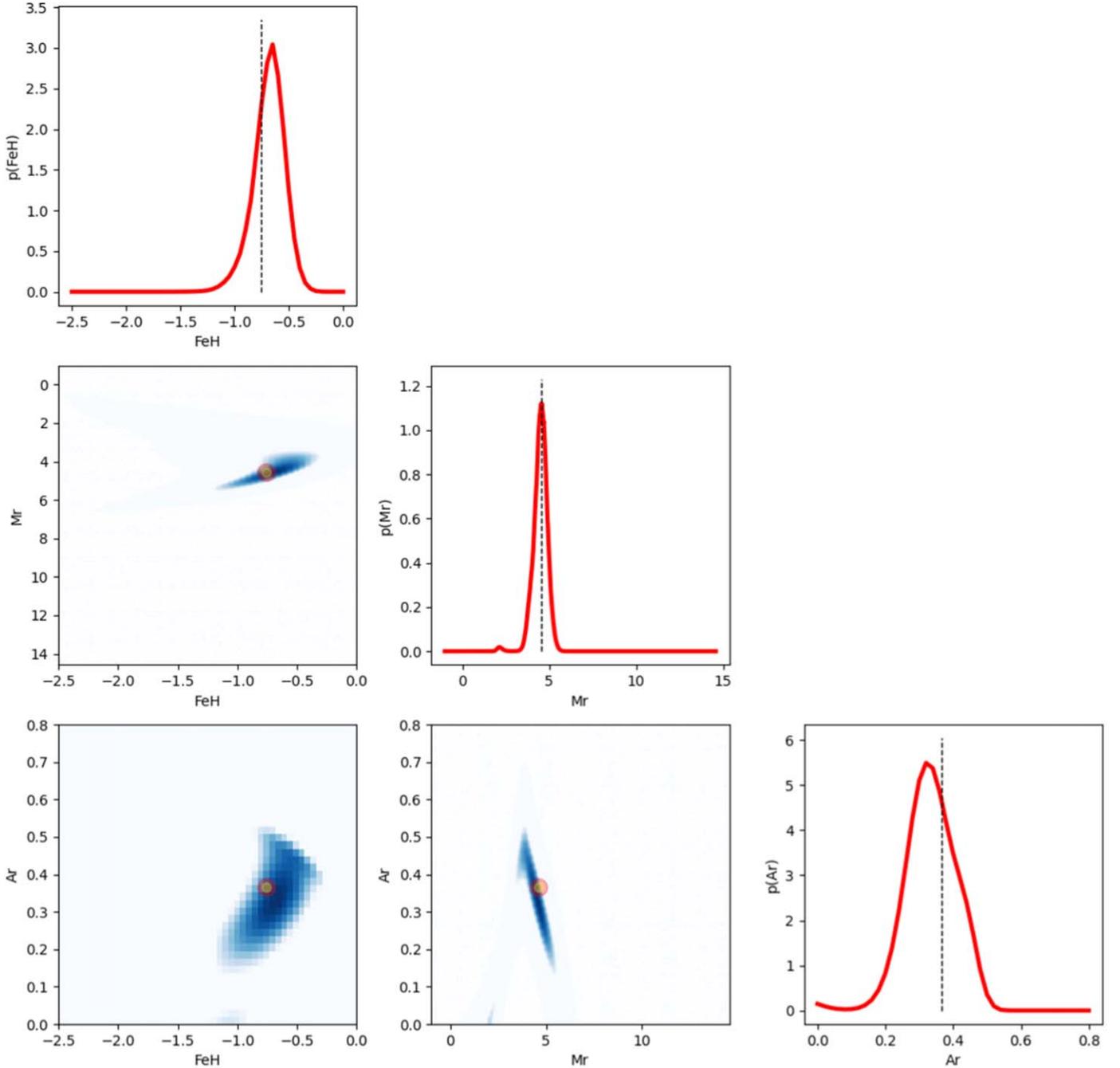

**Figure 9.** Two-parameter covariances and marginal distributions for the posterior map from Figure 8. The symbols and dashed lines show the true input values.

model $k$, $k = 1...N_{\mathcal{M}}$, as

$$p(k) = \frac{O_{k1}}{\sum_{j=1}^{j=N_{\mathcal{M}}} O_{kj}}, \qquad (16)$$

where the model $k = 1$ is chosen arbitrarily without a loss of generality ($O_{11} = 1$).

## 3. Numerical Implementation and Performance Tests

We first discuss our numerical implementation choices and then test the method's performance using simulated TRILE-GAL catalogs. We validate adopted luminosity–color models and implied distance scale using SDSS and Gaia observations. An all-sky pipeline implementation and its performance are discussed in the next section.

### 3.1. Accelerated Exhaustive Grid Search Method

The luminosity–color models introduced in Section 2 are expressed as lookup tables, rather than as analytic functions. They are initially defined on a two-dimensional grid, and after accounting for dust extinction, on a three-dimensional grid. We use equidistant grids for all three model parameters: 0.01 mag for $M_r$, 0.05 dex for [Fe/H], and a range of steps from 0.005 to 0.02 mag for $A_r$, depending on its prior range. These grids have sufficient resolution for numerical analysis of the posterior probability distributions and typically result (depending on the maximum allowed $A_r$ value) in several million grid points. For a given star, the exhaustive grid search method simply iterates over all these points and for each computes the likelihood and ultimately the posterior PDF.

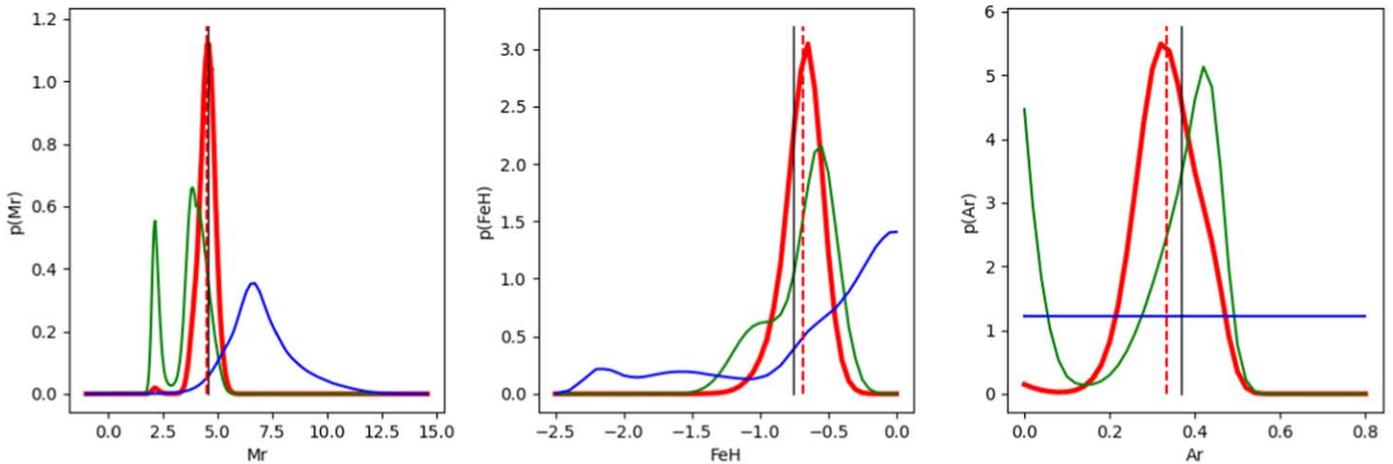

**Figure 10.** An illustration of the improvement in the "knowledge" of model parameters between the prior and posterior, brought by measurements via the likelihood map. The lines show marginal distributions produced with the prior (blue), likelihood (green), and posterior (red) maps from Figure 8. The dashed line shows the expectation value for the marginalized posterior and the vertical solid line marks the true input value.

Significant acceleration can be achieved by using a two-step iteration, with degraded grids in the first step, but still fine enough to sufficiently resolve the posterior PDF to approximately estimate its behavior around its maximum. We find that degrading the $M_r$ step to 0.05 mag and the [Fe/H] step to 0.1 dex works well in practice and results in an about 10 times shorter runtime. After the first iteration, the range $\pm 3\sigma$ around the posterior maximum, where $\sigma$ is standard deviation, is estimated using the marginal distribution for each model parameter. The calculation of the posterior is then repeated using the finest grid but only in the $\pm 3\sigma$ neighborhood (cube) around the posterior maximum, which includes 100–1000 times fewer grid locations.

With this acceleration trick, the runtime per star is typically about 10 ms (it varies nearly proportionally to the maximum allowed $A_r$ value). While exceedingly simple, this method is sufficiently fast and robust to offer a convenient practical solution for processing samples of billions of stars.

### 3.1.1. Markov Chain Monte Carlo Method

Markov Chain Monte Carlo (MCMC) methods have been successfully used in this context (e.g., G. M. Green et al. 2014). Given that we aim to process LSST photometry for about $10^{10}$ stars, it is worthwhile to compare the MCMC runtime to the runtime for our accelerated exhaustive grid search method. We considered two implementations, based on `PyMC3` (A.-P. Oriol 2023) and `emcee` (D. Foreman-Mackey et al. 2013). The latter was about 10–20 times faster than the former; for example, with four "walkers" and 500 iterations, it took on average about 0.1 s per star. Therefore, in this low-dimensional case (there are only three model parameters) the grid search method described above is about an order of magnitude faster than the `emcee` implementation.

### 3.1.2. Neural Network Method

Further acceleration can be achieved using the neural network method. In a companion paper by K. Mrakovčić et al. (2025, in preparation), we describe how to use a small subsample of stars to train a neural network model and then estimate model parameters and their uncertainty for the remaining stars using the trained network. This approach yields runtimes of about 1 ms per star, that is, shorter by an order of magnitude than when the grid search method is used for the full sample. With this additional speed up, a sample of 10 billion stars can be processed in about 10 hr using a 300 core machine and the pipeline described in the next section.

### 3.2. Performance Testing

We first validate numerical implementation of our method using a simulated dataset, where the true model parameters are known. The luminosity–color model tracks are then validated using SDSS photometry from the so-called Stripe 82 region and trigonometric distances obtained using Gaia data. Finally, we demonstrate performance in the large $A_r$ regime and three-dimensional mapping of interstellar dust distribution using SDSS scans that cross high-extinction regions in the Galactic plane.

#### 3.2.1. Simulated Data Set

Our simulated catalogs are based on TRILEGAL simulations of LSST stellar content by P. Dal Tio et al. (2022), already mentioned in the context of priors in Section 2.3. We use these simulations to generate distributions in the $M_r$–[Fe/H]–$A_r$–magnitude–sky position space. Given these quantities for a sample of simulated TRILEGAL stars, we generate simulated LSST photometry for each star as follows.

1. Given relevant model parameters (e.g., $M_r$ and [Fe/H] for main-sequence and red giant stars, $M_r$ and $\log(g)$ for WDs, the total system luminosity, and the component luminosity ratio in the $r$ band for unresolved binaries), we use luminosity–color tracks (see Figure 3) to assign photometric noise-free colors. We do not use the original TRILEGAL colors because for proper statistical tests colors must be consistent with luminosity–color tracks used for likelihood computations.

2. Using the apparent $r$-band magnitude from the TRILEGAL simulation, corresponding to the simulated distance

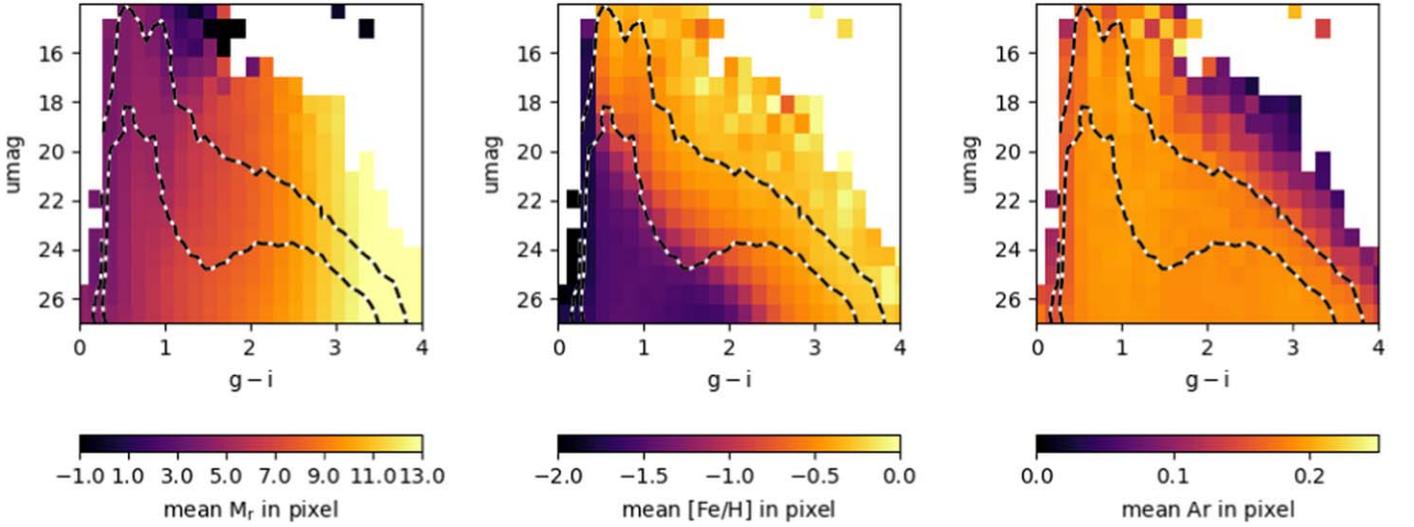

**Figure 11.** The mean per-pixel values of absolute magnitude $M_r$, metallicity [Fe/H], and interstellar dust extinction along the line of sight in the $r$ band, $A_r$, for a TRILEGAL-simulated sample of 280,000 main-sequence and red giants stars with $r < 26$, $u < 27$, $340° <$ R.A. $< 350°$, and $-1°\!.3 < \delta < 1°\!.3$ (a small patch from the SDSS Stripe 82 region) in the $u$ vs. $g - i$ color–magnitude diagram. The mean values are color coded according to the legend below each panel. Contours visualize the sample distribution in each diagram. The strong variation of $M_r$ with the $g - i$ color is seen in the left panel, except for the small patch with $u < 16.5$ and $1.0 < g - i < 1.5$ where red giant stars dominate. The diagonal iso-metallicity boundaries in the middle panel closely correspond to distance (in the range from about 1 kpc to about 100 kpc in the lower left corner). The values of dust extinction, shown in the right panel, are not large (<0.2) because the selected field is at high Galactic latitudes (centered on $b = -52°$).

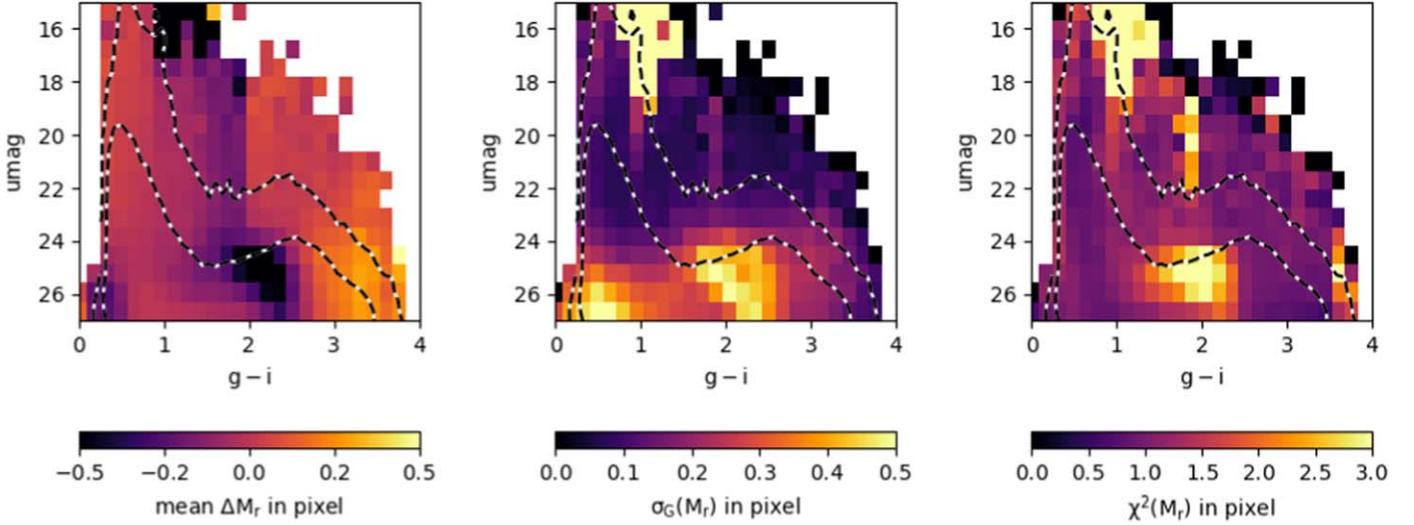

**Figure 12.** Statistical performance analysis for the estimates of model parameter $M_r$ (absolute magnitude), as a function of observed $u$-band magnitude and the $g - i$ color, for the same simulated sample as shown in Figure 11. The left column shows the mean difference per pixel between the true and estimated values, the middle column shows scatter per pixel, and the right column shows the scatter normalized by the estimated uncertainties ($\chi^2$). Contours visualize the sample distribution in each diagram. Note that for main-sequence stars $M_r > 4$ and for most stars $4 < M_r < 10$.

and $M_r$, and colors from the previous step, we generate magnitudes in all the remaining bands.

3. We generate per-band photometric uncertainties, $\sigma_b$, using Equations (4) and (5) from Ž. Ivezić et al. (2019) and per-band 5$\sigma$ depths (the so-called $m_5$) expected for coadded LSST data from F. B. Bianco et al. (2022).
4. We draw Gaussian noise from $N(0, \sigma_b)$ and add it to each magnitude to obtain "observed" magnitudes.
5. Finally, we recompute photometric uncertainties to be treated as "observational" uncertainties using "observed" magnitudes (because the first set of photometric uncertainties are derived using true noise-free magnitudes).

For numerical tests, we select a region from the so-called SDSS Stripe 82, defined by $340° <$ R.A. $< 350°$ and $-1°\!.3 < \delta < 1°\!.3$, which contains 280,000 simulated main-sequence and red giants stars that satisfy $r < 26$ and $u < 27$. Their distributions of model parameters $M_r$, [Fe/H], and interstellar dust extinction along the line of sight in the $r$ band, $A_r$, in the $u$ versus $g - i$ color–magnitude diagram are shown in Figure 11.

Figures 12 and 13 show the summarized statistical performance of the Bayesian method (bias, scatter, and $\chi^2$) in the $u$ versus $g - i$ color–magnitude diagram. In this test based on a high-Galactic-latitude field, we assumed that all stars are beyond the dust layer and fixed $A_r$ to its known true value (in practice provided by dust maps such as those from D. J. Schlegel et al. 1998). When $A_r$ is considered as a free parameter in sky regions with small $A_r$, the best-fit $A_r$ values are often underestimated for stars in the blue part of the stellar locus (because the reddening vector and locus are nearly parallel; for more details, see Section 2.7.1 in M. Berry et al. 2012). We note that at faint magnitudes probed by LSST, this assumption will only break down very close to the Galactic plane.

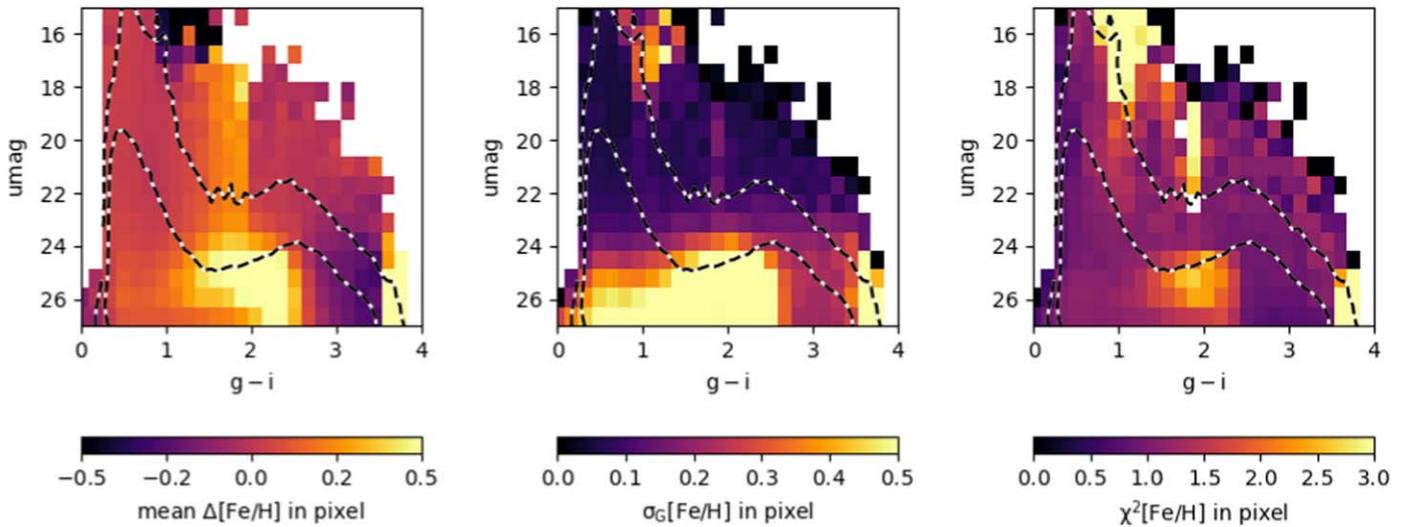

**Figure 13.** Analogous to Figure 12, except for estimates of [Fe/H] (metallicity) instead of $M_r$.

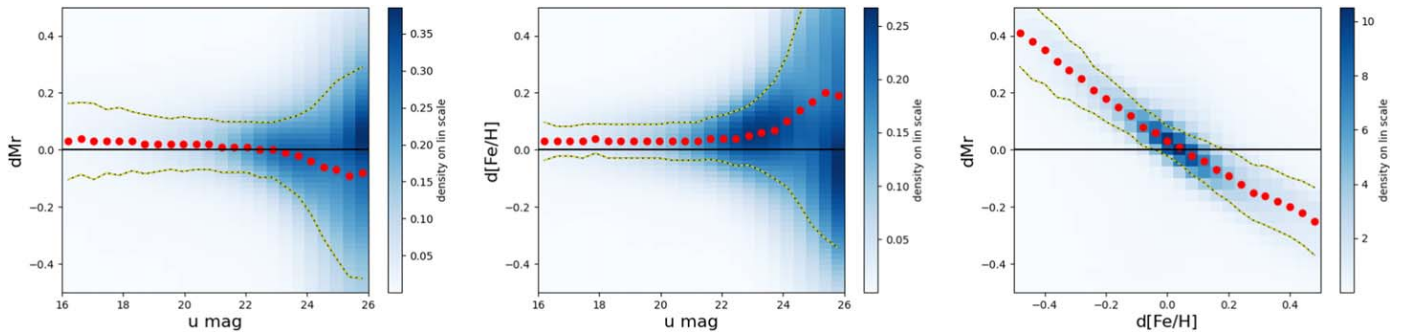

**Figure 14.** The variation of the difference between the true and estimated values (left: $M_r$, middle: [Fe/H]) with $u$-band magnitude. The binned background map shows the same simulated sample as in Figure 11. The symbols show binned medians and the dashed lines show robust standard deviation around the medians. At about $u = 23$, the scatter for both $M_r$ and [Fe/H] starts increasing due to increasing $u$-band measurement error. The dataset was generated assuming a $u$-band $5\sigma$ limiting depth of 25.73 for coadded LSST photometry (F. B. Bianco et al. 2022). The right panel illustrates a strong anticorrelation between the differences.

As discussed in detail in Ž. Ivezić et al. (2008), the photometric metallicity is best constrained for blue main-sequence stars with $g - i < 1$ (equivalently, $0.2 < g - r < 0.6$). Therefore, we expect the best statistical behavior (smallest scatter) in this color range, which is consistent with the behavior seen in Figures 12 and 13. An effective depth limit is about $u = 25$ (about 1 mag brighter than the $5\sigma$ depth in the $u$ band for coadded LSST photometry, F. B. Bianco et al. 2022); at fainter magnitudes the scatter between the true and estimated values for both $M_r$ and [Fe/H] rapidly increases.

Figure 14 shows a more quantitative illustration of the bias and scatter for the best-fit $M_r$ and [Fe/H] as functions of the $u$-band magnitude, which sets the effective signal-to-noise ratio. In the bright limit ($u < 22$; approximately $r < 21$ for blue stars), $M_r$ and [Fe/H] can be estimated with uncertainties of 0.10 mag and 0.07 dex, respectively, and essentially negligible biases (0.02 mag for $M_r$ and 0.03 dex for [Fe/H]). These uncertainties are consistent with simulated photometric errors and intrinsic properties of photometric parallax and photometric metallicity methods (for detailed discussion, see Ž. Ivezić et al. 2008).

This $M_r$ uncertainty implies distance estimates for blue stars accurate to about 5% to a distance limit of 25 kpc, with coadded LSST photometry. Such unprecedented accuracy for photometric distance estimates is due to the ability of the $u$ band to constrain [Fe/H]. For a subsample with $u \sim 24.73$, 1 mag brighter than the $5\sigma$ limit, the scatter increases to 0.27 mag and 0.38 dex, and further to 0.36 mag and 0.54 dex for a subsample at the $5\sigma$ limit in the $u$ band (where the $u$-band photometric uncertainty is 0.2 mag).

As the right panel in Figure 14 demonstrates, estimates of $M_r$ and [Fe/H] are highly anticorrelated (this is essentially a consequence of Equation (A2) from Ž. Ivezić et al. 2008; also visible as the shift of the stellar main sequence with metallicity in the top panels in Figure 3). An [Fe/H] uncertainty of 0.3 dex, approximately the width of the halo metallicity distribution, would induce an additional uncertainty of estimated photometric $M_r$ of 0.3 mag (as would be the case for faint stars with weak or no [Fe/H] constraint coming from their $u$-band measurement).

The bias and scatter for both $M_r$ and [Fe/H] are large in a small patch with $u < 16.5$ and $1.0 < g - i < 1.5$ where red giant stars dominate, due to strong degeneracies in color space between giants and main-sequence stars (recall Figure 4). About two-thirds of all red giants are misidentified as main-sequence stars (due to priors, as they are by and large indistinguishable by colors). Nevertheless, these simulation-based tests suggest that it will be possible to select highly pure samples of red giants using best-fit $M_r$ and a simple $M_r < 3$ cut. This selection criterion selects 28% of all red giants in the sample (which correspond to about 1% of the full sample), with a purity of 99.9%. With LSST data, it will be possible to select such red giant candidates, if they exist, to distances of several hundred kiloparsecs (note that low-[Fe/H] main-sequence stars at the blue edge of the stellar locus and $u = 25$, with LSST distance uncertainties of about 20%, will be detected to distances of about 100 kpc).

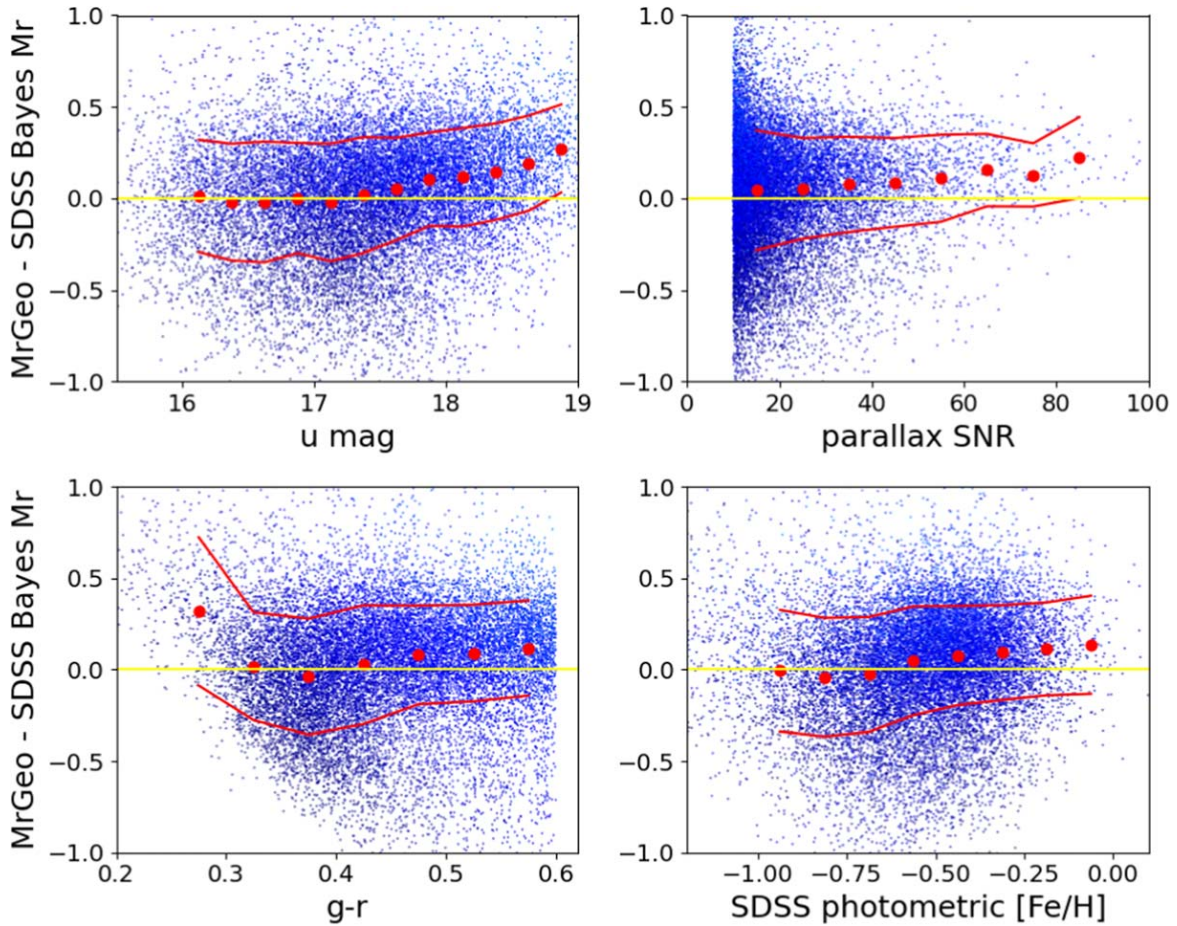

**Figure 15.** Analysis of the difference between Gaia's geometric (trigonometric) absolute magnitudes and the photometric magnitudes derived from SDSS colors. Symbols show a sample of 33,186 stars with a parallax signal-to-noise ratio above 10, absolute magnitude $M_{r,\text{Geo}} > 4$ and $M_{r,\text{Pho}} > 4$, and SDSS-based $u < 21$ and $0.2 < g - r < 0.6$. The median magnitude difference is −0.01 mag, with a scatter of 0.28 mag. The four panels illustrate variations of the binned median (red circles) and scatter (red lines) with the $u$-band magnitude, the parallax signal-to-noise ratio, SDSS $g - r$ color, and photometric metallicity.

### 3.2.2. SDSS-based Comparison to Gaia's Distance Scale

Analysis of the method's performance in the preceding subsection was based on a simulated sample with "observed" colors generated using exactly the same luminosity–color model tracks as those used in fitting. Therefore, that analysis cannot test the validity of those tracks in an absolute sense. While these empirical luminosity–color model tracks were derived using SDSS observations of globular clusters with known distances and metallicities, here we validate them further using trigonometric distances recently obtained by Gaia.

For validation, we used the SDSS Stripe 82 Standard Star Catalog, recalibrated by K. Thanjavur et al. (2021), which provides the most precise available photometry in SDSS bands (due to averaging of many repeated SDSS observations). For stars listed in the catalog, we extracted Gaia measurements and "photogeometric" distances from C. A. L. Bailer-Jones et al. (2021). Out of 841,000 stars listed in the Standard Star Catalog, there are 415,000 stars with $r < 22$, $u < 22$, and a Gaia match within 0.″15 (after correcting for proper motion using Gaia measurements). Their distributions in color–color and color–magnitude diagrams are shown in Figures 1 and 2. We estimated $M_r$ and [Fe/H] for these stars using the same procedure and assumptions as for the simulated sample from the preceding subsection. In particular, we assumed that all stars are beyond the dust layer and fixed $A_r$ to its value provided by the dust maps from D. J. Schlegel et al. (1998).

We first analyze results for a subsample of stars that have $0.2 < g - r < 0.6$, a color range where the photometric metallicity estimator is best constrained, and a signal-to-noise ratio for Gaia's trigonometric parallax measurement above 10. We also required $u < 21$ but that requirement is less stringent than the implied flux limit due to Gaia's signal-to-noise ratio limit (approximately $u < 19$). There are about 18,000 stars in this subsample. We find that their $M_r$ values are estimated using

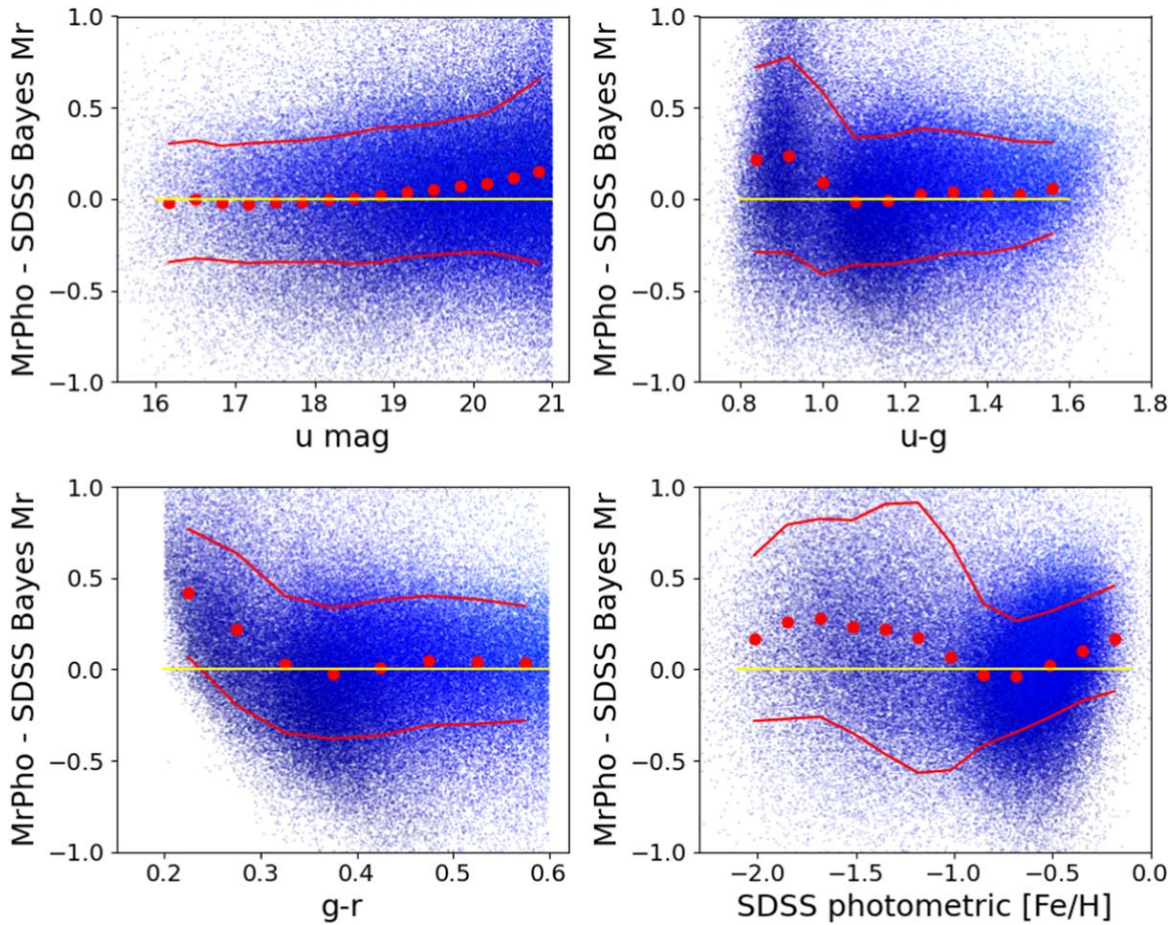

**Figure 16.** Analysis of the difference between Gaia's photometric (color-based) absolute magnitudes and photometric magnitudes derived from SDSS colors. Symbols show a sample of 165,834 stars with absolute magnitude $M_{r,Pho} > 4$, and SDSS-based $u < 21$ and $0.2 < g − r < 0.6$. The median magnitude difference is −0.05 mag, with a scatter of 0.32 mag. The four panels illustrate the variation of the binned median (red circles) and scatter (red lines) with the $u$-band magnitude, SDSS $u − g$ and $g − r$ colors, and photometric metallicity.

SDSS photometry with a bias of −0.06 mag and an rms scatter of 0.30 mag, compared to Gaia's measurements. Equivalently, photometric distances are estimated with a scatter of 15% and a bias of 3%.

The median contribution of Gaia's parallax measurement uncertainty to Gaia's absolute magnitude uncertainty for this subsample is about 0.12 mag. This estimate implies that the median uncertainty of SDSS-based $M_r$ estimates is 0.27 mag (corresponding to about a 14% distance uncertainty). On the other hand, when only stars with a signal-to-noise ratio for Gaia's trigonometric parallax measurement above 50 are considered, the implied uncertainty of the SDSS-based $M_r$ estimates is 0.21 mag (corresponding to about a 10% distance uncertainty). Recalling that the statistical uncertainties for $M_r$ estimates using simulated sample in the preceding section are 0.10 mag at the bright end, these somewhat larger uncertainties may imply additional statistical effects not accounted for in our analysis, and/or imperfect luminosity–color tracks. Figure 15 demonstrates that there are no concerningly large systematic errors in the photometric $M_r$ estimates with respect to photometric noise ($u$-band magnitude), $g − r$ color, and photometric metallicity estimates (though note that the metallicity range corresponds only to disk stars).

We extended our analysis to fainter magnitude limits by replacing Gaia's magnitudes based on trigonometric distance with those based on the so-called "photogeometric" distances from C. A. L. Bailer-Jones et al. (2021). This comparison goes about 2 mag deeper and also extends to a low halo-like metallicity range. The scatter between photometric $M_r$ based on SDSS data and $M_r$ based on "photogeometric" distances is 0.38 mag. As shown in Figure 16, the systematic error increases to about 0.2–0.3 at the blue and low-metallicity edge of the stellar locus. When stars with $0.2 < g − r < 0.3$ are excluded, systematic errors at the low-metallicity ([Fe/H] < −1) end disappear,[18] indicating that it is the $M_r$ as a function of color relation that fails at the blue end, rather than the shift of $M_r$ as a function of [Fe/H]. Such edge effects will have to be recalibrated when LSST photometry becomes available (both using Gaia data and globular clusters).

### 3.2.3. Performance in the Galactic Plane

As the final test, we analyze the method's performance in regions with large interstellar dust extinction ($A_r$). We use

---

[18] The so-called "photogeometric" distances from C. A. L. Bailer-Jones et al. (2021) do not utilize [Fe/H] information. Since the [Fe/H] distribution varies with apparent magnitude, one might expect biased distances (20% or more for halo versus disk comparison). We find no significant bias compared to our estimates, which probably indicates that their distance priors absorb the impact of the unknown [Fe/H]. Hence, the only net effect of the unknown [Fe/H] is increased uncertainty of the estimated distances (for halo stars, with a metallicity scatter of 0.3 dex, the expected contribution to $M_r$ scatter is about 0.2 mag).

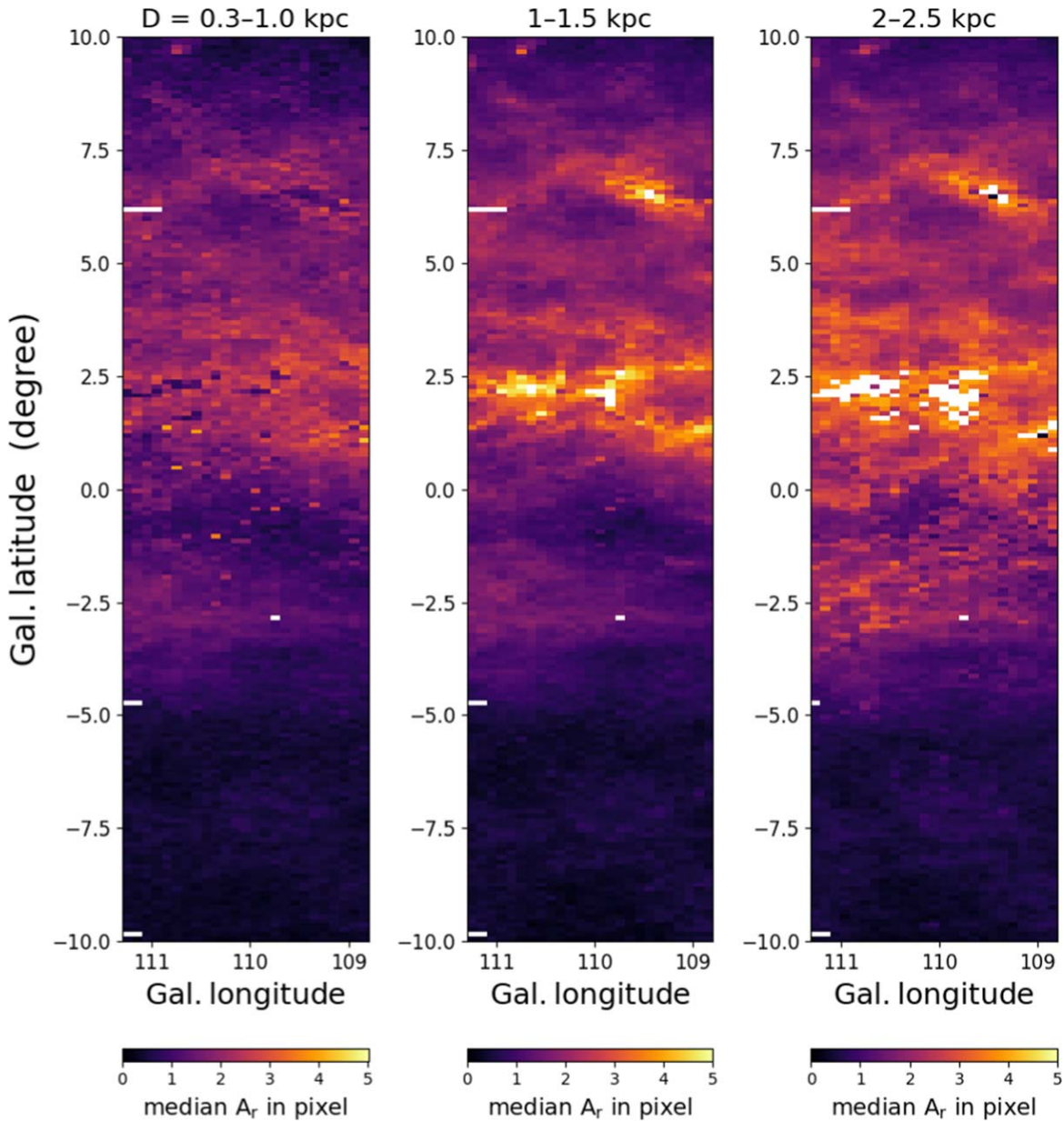

**Figure 17.** Color-coded maps show the best-fit $A_r$ (dust extinction along the line of sight) per pixel for the SDSS-SEGUE strip centered on $l = 110°$ (modeled after Figure 25 from M. Berry et al. 2012). Each pixel, defined in Galactic coordinates, shows the median $A_r$ according to the legend below each panel. The three panels correspond to distance slices of 0.3–1 kpc (left), 1–1.5 kpc (middle), and 2–2.5 kpc (right). Note that the median extinction increases rapidly with distance. In the right panel, there are no stars in the white regions in the center because extinction makes them fainter than the sample magnitude limit.

SDSS-SEGUE data collated and described by M. Berry et al. (2012). For illustration, we use a single strip perpendicular to the Galactic plane and centered on longitude $l = 110°$. Figure 17 shows the best-fit dust extinction along the line of sight for three distance slices. Clear three-dimensional structure of the dust distribution is evident and consistent with Figure 25 from M. Berry et al. (2012).

These tests validate numerical implementation of the Bayesian algorithm and demonstrate its performance on a single patch of data, where priors can be assumed uniform. An all-sky pipeline implementation is described in the next section.

## 4. All-sky PhotoD Pipeline Implementation

Implementation of the Bayesian algorithm described in the preceding sections assumes that the priors can be assumed constant over some sky region, hereafter called a "patch" (see Section 2.3). In practice, patches need to be at least several square degrees large in order to have enough stars to compute priors. At the same time, they should not be too large to satisfy the assumption of constancy. Both of these size limits vary across the sky due to Galactic structure effects.

The scale of the dataset expected from LSST—O(10Bn) objects—makes this computation challenging to perform on a single core or computer. The problem is highly parallelizable and thus, ideally, one would wish to execute the computation in parallel on many (hundreds of) cores, likely housed on a number of separate nodes. This requires a distributed processing framework capable of orchestrating such a solution.

### 4.1. Large Survey Database and HEALPix Implementation

For our work, we have adopted the Large Survey Database (LSDB; S. Wyatt et al. 2024; N. Caplar et al. 2025, in preparation) framework.[19] LSDB is a distributed computing framework written in Python, which enables multinode and multicore computation on large (PB-scale), partitioned, catalog datasets. It extends the concepts of the LSDB toolkit (M. Juric 2011) to astronomical catalogs where the distribution of objects on the sky is very uneven—such as is the case for Galactic star counts. Leveraging broadly adopted community libraries such as astropy, Pandas, and Dask, LSDB presents a user-friendly application programming interface on which it is possible to build our pipeline.

Specifically, LSDB allows us to define a piece of computation (the PhotoD function, written in Python) that should be executed either on every object in the catalog, or on a per-patch basis (where the patch has a potentially dynamically sized solid angle). For example, we compute priors for each of the patches (Section 2.3). LSDB then ensures that these functions are successfully run over the entire (arbitrarily sized) catalog. LSDB handles the data distribution and movement between multiple nodes in the cluster, automatically handles transient failures, and writes the output (in parallel) to another partitioned, searchable, table.

This workflow is particularly efficient and well suited for the scientific objectives addressed by PhotoD, as it leverages the variable resolution of the map based on the stellar density within each region. For instance, in the Galactic plane where the stellar density is higher, smaller pixels provide higher resolution, while in the halo, where the stellar density is lower, larger pixels suffice.

This approach is also computationally advantageous because it maintains consistent chunk sizes for processing. Each data chunk, regardless of its spatial resolution, is approximately the same size, optimizing the execution of the code. In our specific case, the data chunks correspond to pixels that are a few hundred megabytes in size.

### 4.2. Pipeline Performance

We have performed several benchmarks to estimate the performance of the PhotoD code. In our initial tests based on the SDSS Stripe 82 data we found that the distances for a billion stars can be estimated in about 100 days on a Mac M1 Pro laptop using one core (corresponding to 10 ms per star).

In further testing, we used outputs from the TRILEGAL simulation and found that the parallel processing performed with LSDB scaled well with respect to the single-core values. On a local cluster node with dual AMD EPYC 9474F 48 core processors, we were able to achieve per-star figures similar to single-core processing, therefore reducing the required processing time to about a week. Our ongoing tests are showing that when all 10 nodes (400 cores) of the Ruđer Bošković Institute's Narval cluster are used, we may be able to estimate distances for 10 billion stars in about 2 days.

---
[19] https://lsdb.io

The output catalog containing LSST photometry (with errors) and minimalistic auxiliary metadata for 10 billion stars would amount to about 1 TB of data. The code outputs, including a covariance matrix of the three fitted parameters, would have approximately the same volume.

## 5. Discussion and Conclusions

Distances to stars are a crucial ingredient in our quest to better understand the formation and evolution of the Milky Way galaxy. Anticipating photometric catalogs with tens of billions of stars from Rubin's LSST, here we presented PhotoD, a pipeline for computing Bayesian distance estimates that can handle LSST-sized datasets.

The Bayesian model implemented in the PhotoD pipeline builds on previous work (e.g., C. A. L. Bailer-Jones 2011; M. Berry et al. 2012; G. M. Green et al. 2014, 2019; C. A. L. Bailer-Jones et al. 2021) and improves it in several important ways: (i) the use of multiple stellar populations (in addition to the dominant main-sequence stars); (ii) improved color tracks for main-sequence stars and (especially) red giants, including the use of very young (<1 Gyr) populations and an extended [Fe/H] range; and (iii) priors based on sophisticated TRILEGAL simulations (P. Dal Tio et al. 2022) that include multiple stellar populations and account for all principal structural components of the Galaxy.

Extensive, although still preliminary, testing demonstrates the expected statistical behavior of implemented computations and that SDSS-based luminosity–color sequences are supported by more recent direct (trigonometric) distance measurements by Gaia. Tests of pipeline performance show that a sample of $10^{10}$ stars can be processed in a few days using a moderate-size cluster. We intend to process each LSST data release as it becomes available and make outputs accessible via Rubin Science Platform.

We anticipate that the accuracy of resulting distance estimates, as well as estimates of metallicity and interstellar dust extinction along the line of sight, will improve as LSST advances because of the following.

1. As more LSST data are collected, photometric depth will improve as well as photometric calibration.
2. Photometric light curves will improve identification of variable populations such as quasars, RR Lyrae stars and eclipsing binary stars. When proper motion measurements become available, separation of nearby stars from distant stars will improve, too.
3. Luminosity–color sequences will be recalibrated in LSST's photometric system and improved using LSST's own globular cluster data and Gaia's parallax measurements.
4. The extension of photometric coverage to longer IR wavelengths (e.g., using Euclid and Roman Space Telescope survey photometry) will improve constraints on model parameters, especially in Galactic plane regions with large dust extinction (for more details, see M. Berry et al. 2012).
5. Improvements in our understanding of the Milky Way structure will iteratively lead to improvements of Galaxy models such as TRILEGAL, and in turn to improvements of Bayesian priors.
6. In the context of interstellar dust studies, it may be worthwhile to consider hierarchical Bayesian modeling

(e.g., setting priors for the line-of-sight reddening profile, as discussed by G. M. Green et al. 2014).

These improvements will require substantial additional work, but given the implied transformative impact of resulting distance estimates on our understanding of the formation and evolution of the Milky Way, it seems well justified.

### Acknowledgments

This work is financed within the Tenure Track Pilot Programme of the Croatian Science Foundation and the Ecole Polytechnique Fédérale de Lausanne, and the Project TTP-2018-07-1171 "Mining the Variable Sky," with the funds of the Croatian–Swiss Research Programme. Parts were based upon work supported by the National Science Foundation under grant No. AST-2003196.

Ž.I. acknowledges funding by the Fulbright Foundation and thanks the Ruđer Bošković Institute for hospitality. Ž.I., M.J., and N.C. acknowledge support from the DiRAC Institute in the Department of Astronomy at the University of Washington. B.A. acknowledges helpful discussions with Douglas Tucker and the members of the Rubin Observatory Stack Club.

The DiRAC Institute is supported through generous gifts from the Charles and Lisa Simonyi Fund for Arts and Sciences and the Washington Research Foundation. Funding for the SDSS and SDSS-II has been provided by the Alfred P. Sloan Foundation, the Participating Institutions, the National Science Foundation, the U.S. Department of Energy, the National Aeronautics and Space Administration, the Japanese Monbukagakusho, the Max Planck Society, and the Higher Education Funding Council for England. The SDSS website is https://www.sdss.org/.

L.P. acknowledges support from LSST-DA through grant 2024-SFF-LFI-08-Palaversa. This work was conducted as part of a LINCC Frameworks Incubator. LINCC Frameworks is supported by Schmidt Sciences. N.C., S.C., M.D., D.J., K.M., A.I.M., and S.M., are supported by Schmidt Sciences.

This work has made use of data from the European Space Agency (ESA) mission Gaia https://www.cosmos.esa.int/gaia, processed by the Gaia Data Processing and Analysis Consortium (DPAC; https://www.cosmos.esa.int/web/gaia/dpac/consortium). Funding for the DPAC has been provided by national institutions, in particular, the institutions participating in the Gaia Multilateral Agreement.

*Facilities:* Gaia and the Sloan Digital Sky Survey.

*Software:* astropy (Astropy Collaboration et al. 2013, 2018), astroML (J. VanderPlas et al. 2012), HATS (S. Wyatt et al. 2024), Jupyter (T. Kluyver et al. 2016), LSDB (M. Juric 2011; S. Wyatt et al. 2024), matplotlib (J. D. Hunter 2007), numpy (T. E. Oliphant 2006), scipy (E. Jones et al. 2001), seaborn (M. L. Waskom 2021), pandas (W. McKinney 2010), and Python (G. Van Rossum & F. L. Drake 2009).

### Appendix

In this appendix we provide validation of the proper motion systematics and random uncertainties using quasars.

We tested Gaia's proper motions and their uncertainties using spectroscopically confirmed quasars from SDSS Data Release 7. There are ~367,000 SDSS quasars with Gaia nonnegative proper motion errors. Their median proper motion per coordinate is about $0.01$ mas yr$^{-1}$ (indicating no substantial systematic measurement errors) and the median proper motion magnitude is about $1.1$ mas yr$^{-1}$ (indicating a typical measurement uncertainty; the median magnitude of this sample is $G \sim 20$; for the FGKM sample analyzed here, with $G < 18$, the proper motion uncertainties are $<0.15$ mas yr$^{-1}$).

We have verified that the width of proper motion per coordinate normalized by the reported uncertainties (i.e., the width of corresponding $\chi$ distributions) is 1.07 and 1.09, demonstrating Gaia's reliable estimates of measurement uncertainties.

We did not find any significant variation of the median quasar proper motion per coordinate with position on the sky. The only "interesting feature" in the data is increased scatter of proper motion per coordinate measurements in the so-called SDSS Stripe 82 region by about 50% compared to the rest of the SDSS sky. This effect is easily understood as due to the deeper quasar sample in that region (due to details in SDSS spectroscopic target selection) and the increase of Gaia's measurement uncertainties with magnitude (and verified through no substantial increase in the corresponding $\chi$ distributions).

### ORCID iDs

Lovro Palaversa ● https://orcid.org/0000-0003-3710-0331
Željko Ivezić ● https://orcid.org/0000-0001-5250-2633
Neven Caplar ● https://orcid.org/0000-0003-3287-5250
Karlo Mrakovčić ● https://orcid.org/0009-0009-8154-3827
Bob Abel ● https://orcid.org/0000-0001-8418-3083
Oleksandra Razim ● https://orcid.org/0000-0002-3045-0446
Filip Matković ● https://orcid.org/0009-0002-5858-585X
Connor Yablonski ● https://orcid.org/0009-0000-0990-8339
Toni Šarić ● https://orcid.org/0000-0001-8731-8369
Tomislav Jurkić ● https://orcid.org/0000-0002-4993-2939
Sandro Campos ● https://orcid.org/0009-0007-9870-9032
Melissa DeLucchi ● https://orcid.org/0000-0002-1074-2900
Derek Jones ● https://orcid.org/0009-0006-2411-723X
Konstantin Malanchev ● https://orcid.org/0000-0001-7179-7406
Alex I. Malz ● https://orcid.org/0000-0002-8676-1622
Sean McGuire ● https://orcid.org/0009-0005-8764-2608
Mario Jurić ● https://orcid.org/0000-0003-1996-9252